\documentclass[journal]{IEEEtran}

\usepackage{amsmath}
\usepackage[dvipsnames,svgnames,x11names]{xcolor}
\usepackage{multirow}
\usepackage[hidelinks]{hyperref}
\usepackage[flushleft]{threeparttable}
\usepackage[textsize=footnotesize]{todonotes}
\usepackage{comment}
\usepackage[binary-units=true]{siunitx}
\definecolor{BLUE}{HTML}{0000FF}

\usepackage{graphicx}
\usepackage{epsfig}
\usepackage{multirow}
\usepackage{microtype} 
\usepackage{booktabs}  
\usepackage{url}
\usepackage{paralist}
\usepackage{subcaption}
\usepackage[textsize=footnotesize]{todonotes}

\usepackage[linesnumbered,ruled,vlined]{algorithm2e}
\usepackage{dblfloatfix}
\usepackage{algorithmic}
\usepackage{adjustbox}

\usepackage{todonotes}

\usepackage{color}

\graphicspath{{figures/}}
\usepackage{tabularx}
\usepackage{textcomp}
\usepackage{bm}	
\usepackage{amsmath}	
\usepackage{authblk}
\usepackage{makecell} 
\usepackage{pdfpages}
\usepackage{xcolor,colortbl}

\begin{document}




\title{Power Side-Channel Attacks on \\ BNN Accelerators in Remote FPGAs}

\author{Shayan Moini, Shanquan Tian, Daniel Holcomb, Jakub Szefer, and Russell Tessier
\thanks{%
  Shayan Moini, Daniel Holcomb, and Russell Tessier are with the Department of Electrical and Computer Engineering, 
  University of Massachusetts, Amherst, MA 01003 USA (e-mail: smoini@umass.edu; dholcomb@umass.edu; tessier@umass.edu).
  Shanquan Tian and Jakub Szefer are with the Department of Electrical
Engineering, Yale University, New Haven, CT 06511 USA
(email: shanquan.tian@yale.edu; jakub.szefer@yale.edu)}}




\maketitle
\begin{abstract}
To lower cost and increase the utilization of Cloud Field-Programmable Gate Arrays (FPGAs),
researchers have recently been exploring the concept of multi-tenant FPGAs,
where multiple independent users simultaneously share the same remote FPGA.
Despite its benefits, multi-tenancy opens up the possibility of malicious users co-locating
on the same FPGA as a victim user, and extracting sensitive information.
This issue becomes especially serious when the user is running a machine learning
algorithm that is processing sensitive or private information.
To demonstrate the dangers, this paper presents a remote, power-based
side-channel attack on a deep neural network accelerator running in a
variety of Xilinx FPGAs and also on Cloud FPGAs using Amazon Web
Services (AWS) F1
instances. This work in particular shows how to remotely obtain voltage estimates
as a deep neural network inference circuit executes, and how the
information can be used to recover the inputs to the neural network. 
The attack is demonstrated with a binarized convolutional neural network used
to recognize handwriting images from the MNIST handwritten digit database.
With the use of precise time-to-digital converters for remote voltage estimation,
the MNIST inputs can be successfully recovered with a maximum
normalized cross-correlation of 79\% between the input image and the
recovered image on local FPGA boards and 72\% on AWS F1 instances. 
The attack requires no physical access
nor modifications to the FPGA hardware.

\end{abstract}


\IEEEpeerreviewmaketitle

\section{introduction}
\label{sec_intro}

Cloud FPGAs have recently emerged as an important computing 
paradigm where users can rent access to high-end FPGA resources
on-demand from public cloud providers.
Most major cloud providers now offer some form of remote, pay-per-use access to FPGAs
\cite{aws_fpga_2020,alibaba_fpga_2020,xilinx_huawei_2017,baidu_fpga_2020,tencent_fpga_2020}.
Furthermore, recent proposals for multi-tenancy have the promise of increasing FPGA utilization, 
especially in data center settings, by fitting multiple
users' designs onto a single FPGA at the same time.
A number of research projects~\cite{elaraby_virtualizing_2008,
byma_fpgas_2014, chen_enabling_2014, weerasinghe_enabling_2015,
khawja_sharing_2018, vaishnav_survey_2018} have explored how to implement FPGA multi-tenancy.
The sharing of an FPGA by many users, unfortunately, opens up multi-tenant FPGA platforms
to many new, potential attacks in which a malicious user
can be co-located next to a victim user.

Once co-located, a malicious user can try to learn information
about the victim through a side channel.  When multi-tenant FPGAs
are deployed in a remote data center, the malicious user is limited to
only using side channels that do not require physical access. 
For example, previous work \cite{giechaskiel_leaky_2018, giechaskiel_leakier_2019,
giechaskiel_measuring_2019, ramesh_fpga_2018,
provelengios_characterization_2019} has shown that crosstalk between long routing wires
on an FPGA can be used to leak sensitive information from cryptographic circuits using remote attacks.
{\color{black} Meanwhile, voltage and power-based attacks \cite{glamocanin:date20} have been used to remotely extract encryption keys for 
both RSA \cite{zhao} and AES \cite{schellenberg:date18} using circuits
implemented on an FPGA by a malicious user.}

The danger of such attacks becomes especially worrisome as there is more and
more interest in the FPGA acceleration of machine learning 
for image
recognition, or other tasks, where sensitive information is processed. 
Existing work on machine learning (ML) algorithm accelerators, 
and especially deep neural networks, using FPGAs~\cite{zhao_fpga_2017, chen_fpga_2019, zeng_fccm_2020, moini}
has shown 
that these algorithms, when deployed on FPGAs, can significantly speed
up the inference operations.
Further, many cloud providers tout FPGAs for acceleration of ML
workloads~\cite{microsoft_ml_2020}.

To show potential threats when machine learning accelerators
are combined with multi-tenant FPGA deployment,  
this work demonstrates a remote power-based side-channel
attack on a binarized convolutional neural network (BNN) in an FPGA.
In our attack, voltage fluctuations, caused by the changes in the power consumption
of the convolution unit in the BNN, are used to accurately reconstruct
images that are input into the BNN accelerator during the inference operations.
Being able to recover the images that are processed by the ML algorithm could reveal
sensitive imagery \cite{wei2018know}. 
To highlight the dangers of the potential attacks, this work shows how to recover such input images remotely, where an attacker uses a time-to-digital (TDC) converter as a remote power sensor in a multi-tenant FPGA setting. Outside of multi-tenant scenarios, the same attack could be used whenever the ML accelerator resides on an FPGA alongside untrusted 3rd-party intellectual property (IP) cores that might contain unknown sensing circuits.

Our attack can be performed 
{\em remotely with no physical hardware access by the attacker}.  
{\color{black} Furthermore, the attack works without knowledge of the neural network
parameters that could facilitate attacks involving power
dissipation templates.}
We demonstrate the details of our attack on the convolution unit of a BNN-based 
circuit that is used to recognize the handwriting images from the MNIST handwritten digit database
\cite{lecun2010mnist}. Our attack and
corresponding image recovery is successfully demonstrated on
multiple generations of Xilinx FPGAs including 
a ChipWhisperer CW305 board~\cite{chipwhisperer} (Artix-7), 
a ZCU104 board~\cite{ZCU104} (Zynq UltraScale+),
a VCU118 board~\cite{VCU118} (Virtex UltraScale+),
and Amazon AWS F1 instances~\cite{AWS} (Virtex UltraScale+).
Based on the evaluation we show that clearly recognizable images can be
retrieved for all tested input images from the MNIST database. A maximum
cross-correlation of 79\% is observed between the original and
recovered images on local FPGA boards and 72\% on AWS F1 instances.

{\color{black}
In summary, our work makes the following contributions:

\begin{itemize}

\item We demonstrate a side channel attack on an ML
accelerator implemented in remote FPGAs. Input images to the
accelerator are reconstructed using TDCs that are logically isolated
from the accelerator.
\item Our attack is shown to work effectively on cloud FPGAs that are
part of AWS F1 instances.
\item We characterize the effectiveness of the attack using
quantitative metrics and examine its robustness to noise. 

\end{itemize}
}

\subsection{Paper Organization}
The remainder of this manuscript is organized as follows.
Section \ref{sec_background} provides 
background on deep neural networks and existing attacks. 
Section \ref{sec_detailed} gives details of our attack and our
experimental approach is described in Section \ref{sec_exp_approach}.
Attack characterization with a ChipWhisperer board is described in Section \ref{sec_experiments}.
Image extraction results generated from commodity FPGA boards and AWS F1 instances are presented in Section \ref{image_extract}. 
Section \ref{sec_conclustion} concludes the manuscript and offers directions for future work.

\section{background and related work}
\label{sec_background}
In this section, we provide an overview of convolutional neural network
models and review previous attacks against
FPGA accelerators for deep neural networks.

\subsection{Convolutional Neural Networks}

Deep neural networks (DNNs) \cite{sze_survey_2017} are a class of artificial neural networks
that use multiple layers.
In a DNN, each layer is
responsible for extracting relevant features, and the output of each layer
is passed as the input to the next layer. 
DNNs combine feature extraction with the classification
capability of classical neural networks to map input data to a set of predictions. DNNs can be used to perform, for example, image classification tasks.

\begin{figure}[t]
	\centering
	\includegraphics[width=0.98\linewidth]{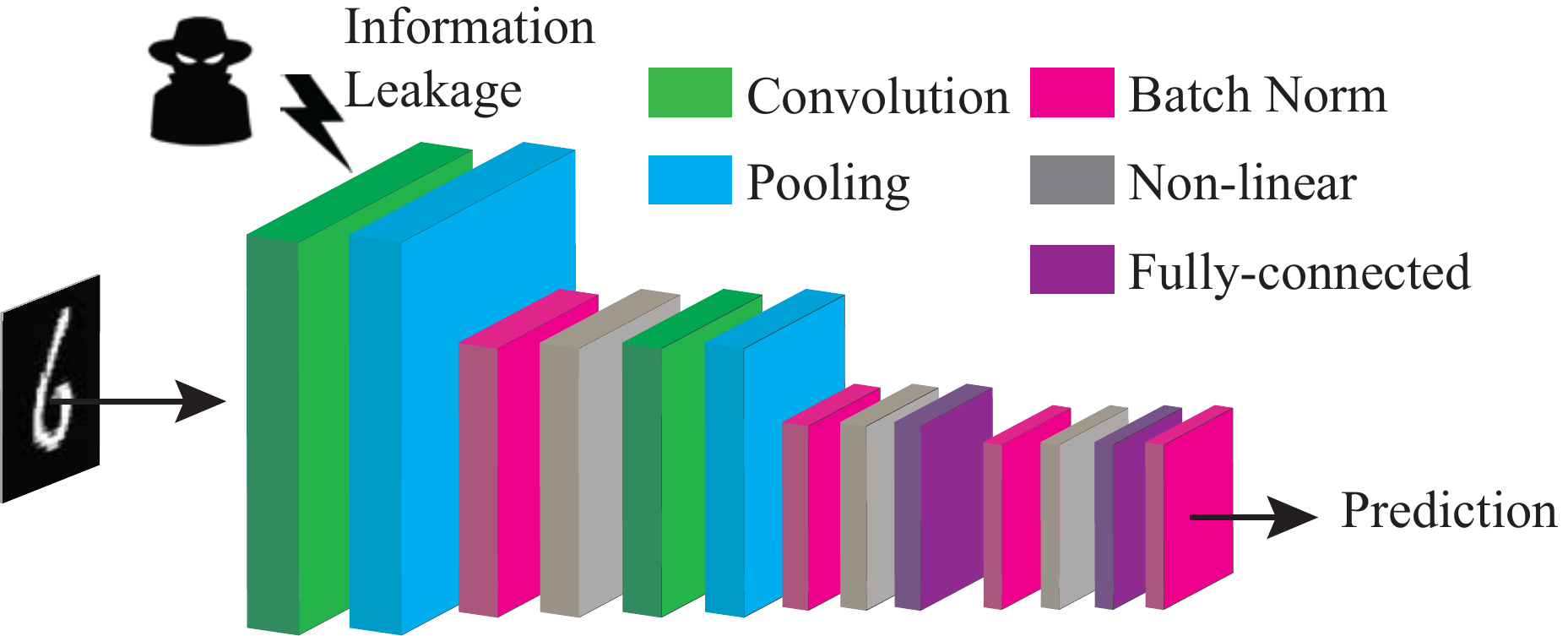}
	\caption{\small Overview of steps in the CNN used in this work. The details of the architecture are explained in Table~\ref{table:1}, and the threat model is shown in Figure~\ref{fig:threat_model}.}
	\label{fig:cnn_overview}
\end{figure}

Convolutional neural networks (CNNs)~\cite{sze_survey_2017} are a subset of DNNs that are
mostly used for classifying multi-dimensional data (e.g., images or
video). 
The main distinctive property of CNNs is the convolution layer,
which implements feature extraction by performing a convolution
operation between the high-dimensional input data (called input feature
maps) and kernels (small matrices of parameters that are computed
during the training phase) to generate the output of the layer (called
output feature maps). 
As shown in Figure~\ref{fig:cnn_overview}, the output feature maps of each layer are passed to
the next layer as the input feature maps. Other layers
in a typical CNN include a non-linear function (creating complex input-output mappings), pooling (reducing
the dimensionality of input feature maps by different methods, e.g.,
max pooling), batch normalization (normalizing input feature maps to
decrease their variance), and fully-connected layers (where each
element of an output feature map is calculated by point-wise multiplication between a whole input feature map and a kernel of the
same size).

\subsection{Binarized Neural Networks}

Binarized neural networks (BNNs)~\cite{binarized} use aggressive quantization
so that each element of the convolution kernel
can be represented as either $-1$ or $+1$. This quantization helps reduce
the memory bandwidth needed to load network parameters from off-chip
memory during the execution of each layer and replaces multipliers with simple add and subtract operations. 
In BNNs, all convolution input feature maps and
kernels are comprised of binary values except for the first input layer which generally
receives its input feature maps as matrices of integers, e.g., representing the pixels of input images.

For this work, we assume the input to the BNN is a grayscale
image with each pixel being represented by an integer (0 to 255).
This image is the input feature map to
the first convolution layer which convolves the input with $n \times n$
binary kernels. To perform the convolution, each element of the 
convolution output (an output feature map) is calculated by
multiplying a kernel with a $n \times n$ window of the input feature map and
summing the resulting values. Sweeping an $n \times n$ kernel across
the input feature map generates an output feature map. The convolution
operation is followed by a maximum pooling operation which reduces the
size of its input feature maps by choosing the largest value out
of each $k \times k$ window of each input feature map and discarding the rest.
The next layer, batch normalization, normalizes its input feature maps
value by value. Here, the numbers are represented as 
fixed-point
values between -1 and +1. The non-linear function layer truncates the
output feature map values of the batch normalization layer into either
-1 or +1 based on their sign. 
This process is replicated for other convolution steps with the
exception that their input feature maps are the binary outputs of the
previous non-linear function layer. 

For this work, the BNN is pre-trained with the MNIST database
on an Nvidia GTX 1080 GPU, 
and the derived parameters, including convolution kernel values,
are used in the BNN accelerator on an FPGA.
We used the Keras framework \cite{ke-ding} to train the BNN.
The trained network is used during the inference stage to classify the input
images of digits into one of ten categories (0 to 9). The BNN contains two
convolution layers and two fully connected layers. 
Convolution is performed with a standard 64 kernels per layer \cite{ke-ding}.
All convolution and
fully-connected layers, except for the first layer, receive binary
inputs, have $3\times3$ binary kernels (e.g., $n=3$), and generate output feature maps in
integer format. The first convolution layer receives the input image, a
28$\times$28 pixel grayscale image of a handwritten digit, as a matrix of integer values
between 0 and 255 and performs the convolution operation with binary
kernels. 
The output of the network is a ten element array that shows the
likelihood of the input image being each of the ten digits with
the highest number being the predicted digit for the input image. Table~\ref{table:1} shows the details of the BNN
architecture used in this work.

\begin{table}[t]
\caption{\small Details of the trained BNN. The accuracy of the trained network
with the MNIST test set is 98.24\%.} 
\label{table:1}
\centering
{\small
\begin{tabular}{|l | l | r | r|} 
 \hline
 Layer \# & Layer Type & Input Size & Kernel Size \\ [0.5ex] 
 \hline\hline
 1 & Convolution & 28$\times$28 & 3$\times$3$\times$64 \\ 
 2 & Pooling & 28$\times$28$\times$64 & 2$\times$2 \\
 3 & Batch norm  & 14$\times$14$\times$64 & - \\
 4 & Non-linear function & 14$\times$14$\times$64 & - \\
 5 & Convolution & 14$\times$14$\times$64 & (3$\times$3$\times$64)$\times$64 \\
 6 & Pooling & 14$\times$14$\times$64 & 2$\times$2 \\
 7 & Batch norm & 7$\times$7$\times$64 & - \\
 8 & Non-linear function & 7$\times$7$\times$64 & - \\
 9 & Fully-connected & 7$\times$7$\times$64 & 500 $\times$ (7$\times$7$\times$64) \\
 10 & Batch norm & 500 & - \\
 11 & Non-linear function  & 500 & - \\
 12 & Fully-connected & 500 & 10$\times$500 \\
 13 & Batch norm & 10 & - \\
 \hline
\end {tabular}
}
\end{table}

\subsection{Attacks on DNN FPGA Accelerators}

Several researchers
\cite{wei2018know,dubey2020maskednet,hua2018reverse,Electromagnetic} have
explored side-channel attacks on DNN
accelerators on FPGAs. All of these approaches used physical access to
the FPGA to collect needed information for the attacks. Meanwhile, we present
a remote, power-based side channel that does not require
physical access to FPGA supply voltage pins, uses on-chip voltage
sensors to detect voltage fluctuations, and is demonstrated to work
with four different FPGA boards.

Wei et al.~\cite{wei2018know} used power traces recovered from FPGA
voltage supply pins to extract the input image data of a BNN.
An oscilloscope was
used to measure the voltage drop across a 1\( \Omega \) resistor placed
on the power supply rail of a SAKURA-G board \cite{SAKURA-G}. 
Their attack method relies on per-clock cycle power
consumption of convolution operations.
Dubey et al. \cite{dubey2020maskednet} targeted an FPGA accelerator of a fully-connected BNN.
They were able to successfully extract the parameter values of the model by finding the 
highest correlation of the model power consumption for a collection of known input values. Voltage traces gathered by an oscilloscope connected to the supply voltage pin of a Kintex-7 FPGA on a SAKURA-X board \cite{SAKURA-X} were used to perform this attack.

Yoshida et al.~\cite{Electromagnetic} used FPGA side-channel
electromagnetic leakage measurements 
to extract the kernel values of a multi-layer perceptron (MLP) 
accelerator in the presence of weight encryption. An external probe was
used to collect these measurements.
Hua et al.~\cite{hua2018reverse} extracted the structure 
of a CNN, 
including the size of the input feature map and kernels of each layer by studying the off-chip memory access 
patterns of the FPGA accelerator while the operations of each layer were performed. Their attack revealed the structure of neural networks in the presence of weight encryption. 
{\color{black} However, they did not reverse 
engineer the input feature map values.} 

Boutros et al. \cite{boutros_2020} recently performed a fault-injection
attack on a CNN implemented in a remote Intel
Stratix 10 FPGA.
Their experiments showed that the deliberate use of excessive power
consumption on the FPGA was not sufficient to cause classification
errors in the CNN due to large timing margins in the circuit
implementation and redundancy in the CNN model.

\subsection{Voltage Sensing Using TDCs}

In FPGAs, small drops in supply voltage occur in the vicinity of power
consumption due to both resistive and inductive drops in the power
distribution network and the chip packaging~\cite{Gnad}. Given that the
propagation delay of combinational logic varies as a function of supply
voltage, circuit delay in a specially designed sensor circuit can be used as a proxy for
measuring the changes in the supply voltage. This approach is commonly
used in voltage sensors based on ring oscillator (RO)
\cite{zick2013sensing} or
TDC \cite{schellenberg:date18} circuits. 
ROs need long measurement periods for precision and are therefore
unsuitable for side channels that rely on fast transients.  
{\color{black} Meanwhile, TDCs are often used to overcome the
limitations of ring oscillator-based sensors~\cite{zick2013sensing} and
have been shown to effectively obtain side channel information on FPGAs~\cite{schellenberg:date18}}. In
TDCs, each measurement reflects the delay of a
circuit within a single clock cycle by observing how far through a
tapped delay line a signal can travel during the cycle. This makes TDC
sensors suitable for sensing short transient voltage fluctuations on
the order of a single clock cycle. Because delay changes are only
observable if they cause the signal to reach the next tap in the delay
line, the precision of a TDC is limited by the delay between successive
taps. As we show in Section~\ref{detailed-tdc}, following others who
previously exploited TDC
designs~\cite{schellenberg:date18,zick2013sensing}, the
high-speed carry logic in modern FPGAs makes a suitable delay line with
taps that are on the order of 5-25 picoseconds (ps) apart, depending on the FPGA
technology and architecture.

{\color{black}
This manuscript significantly extends an earlier work that
examined FPGA image extraction from a BNN model using TDCs \cite{moini:date21}.
We comprehensively explore the issue of TDC-based
image extraction from BNNs in FPGAs by applying denoising to recovered images and 
deliberately stressing the FPGA power distribution network. Unlike
earlier work, our attack is applied to AWS F1 platforms, a commercial
cloud FPGA environment. 
}

\section{details of the attack}
\label{sec_detailed}
In this section, we provide an overview of our threat model.
We then focus on the details of the attack and its implementation in a
multi-tenant FPGA setting.

\subsection{Threat Model}

This work focuses on a multi-tenant FPGA scenario where the victim
user is running a machine learning inference algorithm on a hardware
module that is co-located on the same FPGA with the malicious user's modules. The adversary simultaneously uses the FPGA platform
without sharing logic or I/O resources with the victim.
The victim circuit's input (e.g., the input image) is
sent to the BNN accelerator in the FPGA in a secure manner (e.g., the input may be encrypted). The same input image
is sent by the victim to the FPGA multiple times, a common case in video processing (e.g., of surveillance images). It is assumed that the
adversary is not able to access the inputs. Hence the goal of the
adversary is to learn the inputs. Further, the adversary is not able to
learn the inputs through information leakage (e.g., crosstalk) on the
input wires, which would make the attack trivial. The output is likewise assumed to be
securely sent back to the user, and the adversary is not able to learn the
output directly (if they did, they again would not need the attack).

In this work the focus is on using a TDC to measure voltage changes.  
The data from the TDC is used by the adversary
to estimate the voltage
drop across the FPGA power distribution network (PDN) during the
execution of the convolution layer, as the BNN accelerator does the image classification.
The acquired voltage estimates serve as a side channel 
that can be used to extract the victim's input image data. 
The recovered image approximates the input image by distinguishing
between foreground and background pixels of the image. 

\subsection{Attack Implementation}
\label{detailed-tdc}
\begin{figure}[t]
	\centering
	\includegraphics[width=0.48\textwidth]{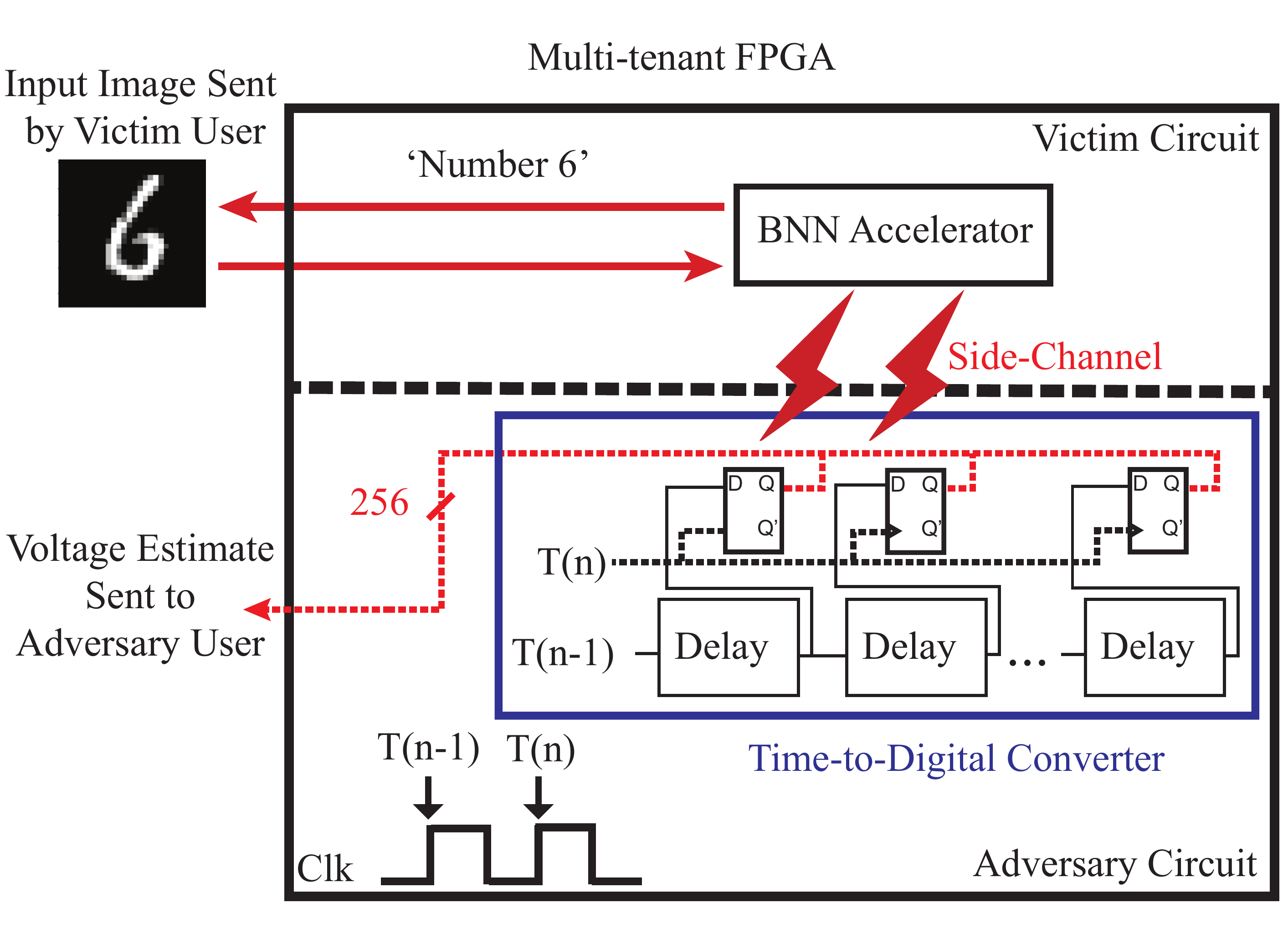}
	\caption{\small Overview of attack implementation. The TDC outputs voltage estimates for each clock cycle of the first convolution layer. These estimates are used to reconstruct the input image.}
	\label{fig:threat_model}
\end{figure}

The attack implementation details are shown in Figure \ref{fig:threat_model}.  In this setup, there is a victim circuit and an attacker circuit co-located on same FPGA. 
To extract the input image from the BNN accelerator, the adversary
focuses on the first convolution layer which directly processes the
input image. The TDC outputs voltage estimates
during each clock cycle of the interval when the
BNN accelerator processes the first convolution layer. The estimates are measured 
using the TDC sensor.

\begin{figure}
	\centering
	\includegraphics[trim={0.5cm 0.1cm 0.1cm 0.5cm},clip,width=0.48\textwidth] {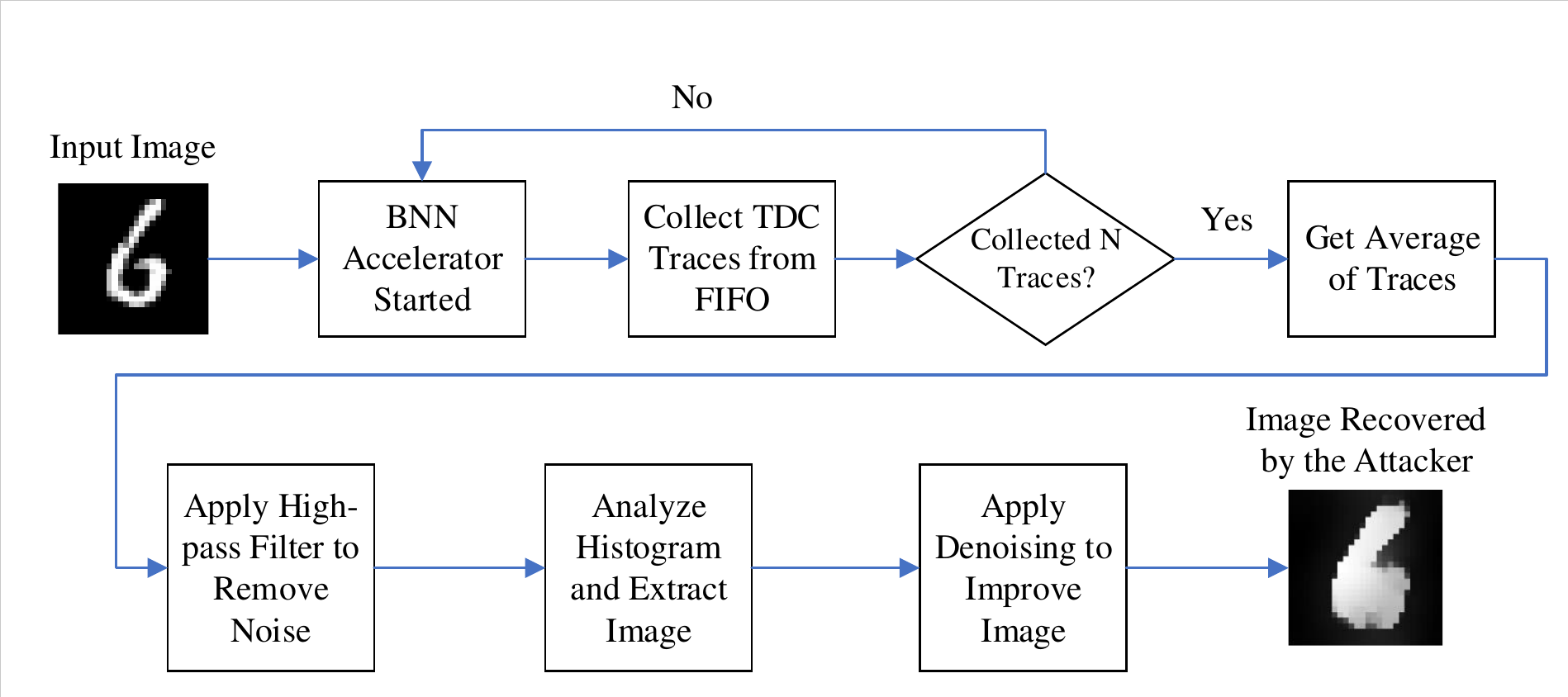}
	\caption{\small Steps of the attack.}
	\label{fig:detailed_operation}
\end{figure}

In the first convolution layer, an image is convolved with multiple distinct
kernels to generate multiple output feature maps. In our attack, we
use a voltage estimate trace from the execution of the first kernel of the first
convolution layer for an input image. Since we assume that the same image is evaluated
by the FPGA accelerator multiple times, multiple ($N$) similar traces are
collected using the same input image. 
After collecting multiple traces, the adversary takes the mean of the data values in the traces 
to obtain a single average trace of the voltage estimates during the
execution of the first kernel of the first convolution layer. 
A high-pass filter is then used to remove noise.
We leverage the observation that the
background and foreground pixels can then be distinguished by analyzing the different magnitudes
of the voltage in a trace of measurements. This information can be represented by a histogram of instance counts of magnitude values in the filtered trace. Points in the histogram are used to label pixels as belonging to the image foreground or background based on the magnitude of their voltage measurement. 
An image denoising filter is applied to this preliminary recovered image to improve clarity. %
The result of the analysis is a reconstructed image that approximates the
input image that was input to the BNN. The procedure is shown in
Figure~\ref{fig:detailed_operation} and discussed in more detail in
Section~\ref{sec_experiments}.

The convolution operation can be represented as~\cite{wei2018know}: 

\begin{equation}
O_{x, y}^{j}=\sum_{i=1}^{M}\left(\sum_{a=0}^{K_{x}-1}
\sum_{b=0}^{K_{y}-1} \omega_{a, b}^{i, j} \times I_{x S_{x}+a, y
S_{y}+b}^{i}\right)\label{eq:1}
\end{equation}

The $O_{x, y}^{j}$ parameter represents the location (x,y) in the
$j$th output feature map which is calculated by convolving a window
(same size as the kernel) of the $i$th input feature map ($I^{i}$) and
the corresponding kernel ($\omega_{a, b}^{i, j}$) and then adding the
$M$ results together where $M$ equals the number of input feature
maps. The $S_{x}$ and $S_{y}$ values represent the convolution step
sizes which are equal to 1 in our BNN implementation. The $K_{x}$ and
$K_{y}$ values represent kernel sizes in the $x$ and $y$ dimensions.

\begin{figure}[t]
	\centering
	\includegraphics[width=0.48\textwidth]{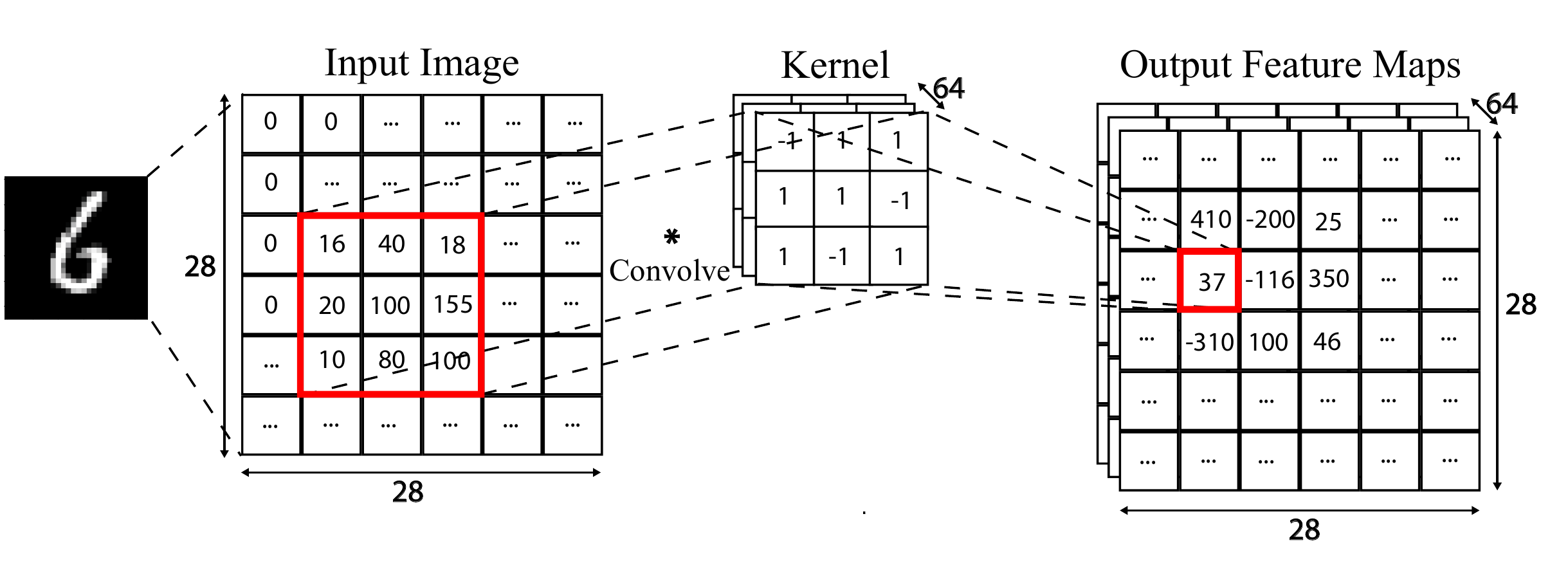}
	\caption{\small Detailed view of the first convolution layer in BNN. The value of 37 shown in the output feature map is generated from the 3$\times$3 input image on the left and the kernel.}
	\label{fig:bnn_overview}
\end{figure}

\begin{figure}[t]
	\centering
	\includegraphics[width=0.48\textwidth]{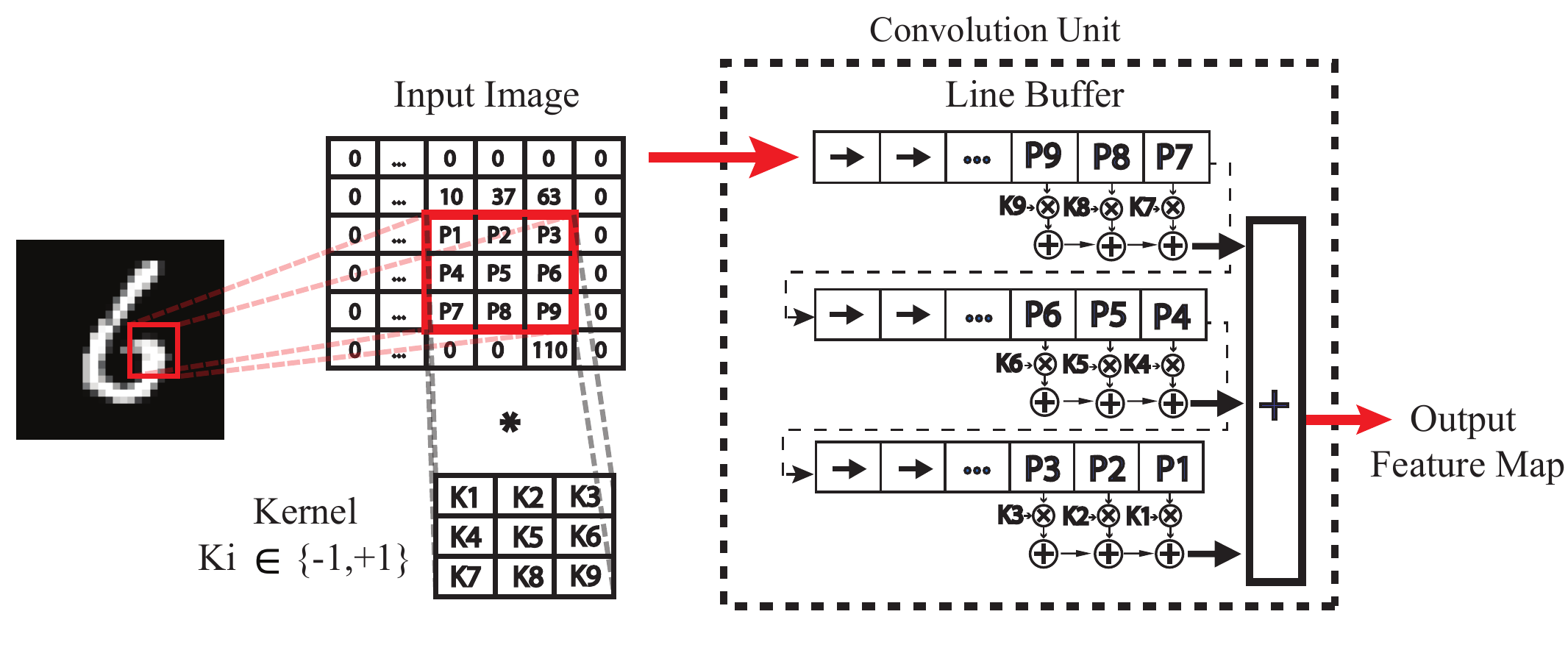}
	\caption{\small Detailed view of the convolution unit. Adapted from~\cite{wei2018know}. Output is generated from the 3$\times$3 input image, shown in the red box, and the kernel. 
	}
	\label{fig:detailed_conv_op}
\end{figure}

For the first convolution layer
of a BNN trained on the MNIST handwritten digit database
with a 28$\times$28 grayscale image as the input and 64 kernels of size
3$\times$3, Equation~\ref{eq:1} can be simplified to:

\begin{equation}
O_{x, y}^{j}=\sum_{a=0}^{2} \sum_{b=0}^{2}
\omega_{a, b}^{j} \times I_{x + a, y + b}, ~~~j \in
[1,64]\label{eq:2} 
\end{equation}

\noindent and represented by the operations shown in Figure \ref{fig:bnn_overview}.

The convolution unit uses a line buffer architecture to hold
and provide data values to the convolution~\cite{wei2018know}. 
As shown by the line buffer at the right in Figure~\ref{fig:detailed_conv_op}, the line buffer is
arranged in three rows, each of 
which processes one line of the convolution operation. 
The line buffer is a shift register that receives one pixel from
the input feature map (the image) per clock cycle and shifts its values
to the right. 
The length of each row in the line buffer matches the
length of the input feature map of the convolution operation (28 for
the first layer in our implementation). The rightmost word of each row
of the line buffer enters the next row from the left, and the rightmost
word of the last row is discarded. The rightmost three words of each of the three
rows of the line buffer constitute the image window whose values are
multiplied with values from the $3\times$3 kernel. Since binary kernels
are used in a BNN, each image
pixel in the current image window is added to or subtracted from (based
on a kernel value of +1 or -1) the other pixels in one clock cycle
using a combinational adder tree. 
One output feature map value is generated every clock
cycle.

\begin{figure}
	\centering
	\includegraphics[width=0.48\textwidth]{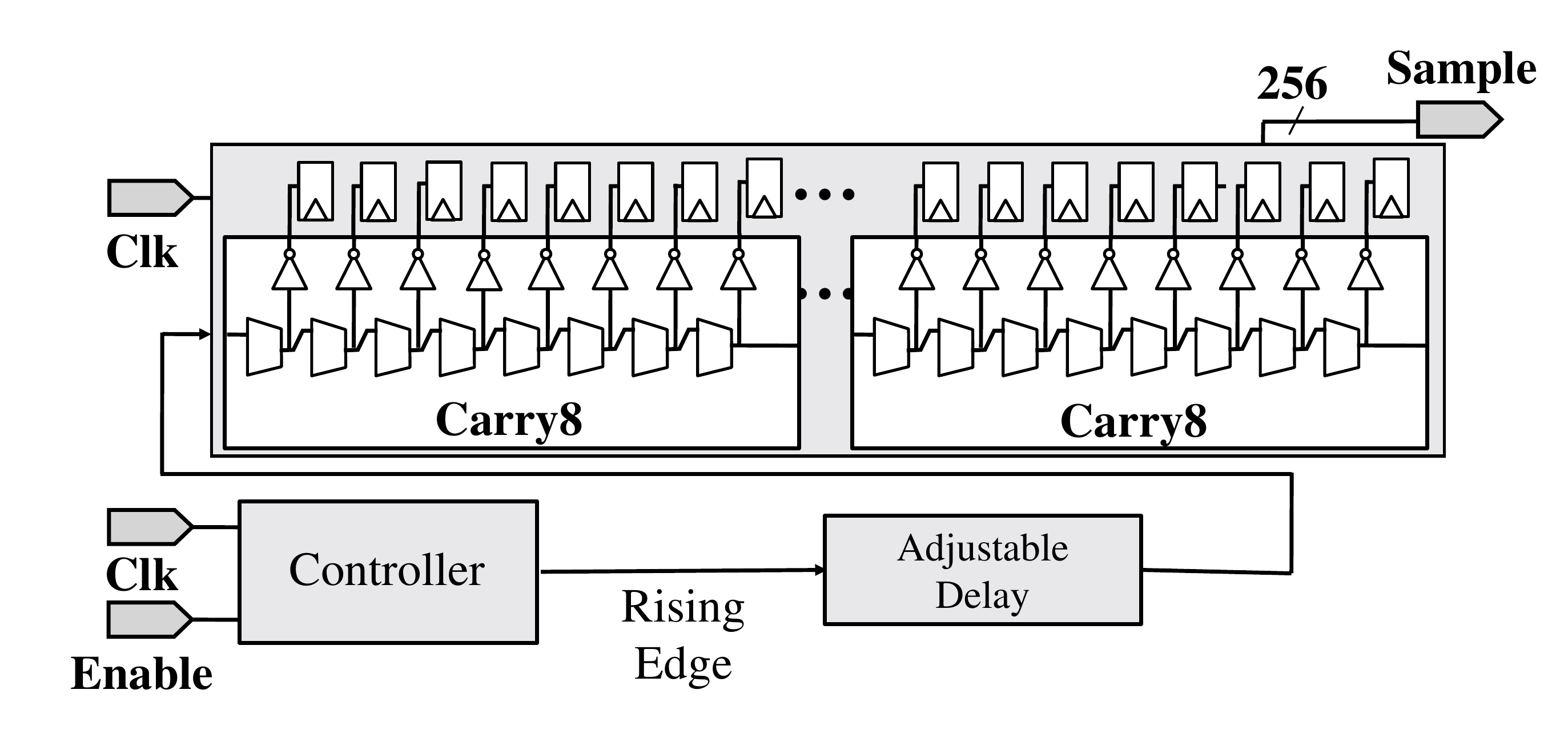}
	\caption{\small {\color{black} Architecture of the TDC.}}
	\label{fig:tdc_architecture}
\end{figure}

An adversary can take advantage of the shared FPGA PDN to sense local
supply voltage changes, which can reveal information about the per-cycle power
consumption in the convolution unit. The power consumption is due in
part to the switching activity in the BNN accelerator, including the
convolution unit, which causes supply voltage to be correlated to the
data processed (larger magnitude data values lead to increased switching). The small PDN fluctuations are reflected in the sampled
values of the time-to-digital converter (TDC), and the TDC samples are
then used to recover a facsimile of the input image. 
The 256-stage TDC architecture is shown in Figure~\ref{fig:tdc_architecture}.
The 256-stage TDC contains an adjustable delay followed by a chain of fast fixed-purpose
FPGA elements typically used to perform timing-critical carry
operations in arithmetic circuits ({\em Carry4} or {\em Carry8} depending on FPGA family). TDC elements are manually placed in
the FPGA for controlled and predictable delay that is matched to the
clock frequency at which the TDC operates.

The TDC is activated by sending the rising edge of a clock through the adjustable delay
and the carry chain to the flip-flops attached to the 256
stages of the carry logic. 
The Hamming weight of the sample indicates how far through
the carry chain the rising edge has propagated by the time the next rising clock edge
arrives. When the supply voltage drops, the propagation delay of the
circuit increases, and the rising edge will have propagated through
fewer carry stages before the next rising clock edge, and hence the sample captured in the flip flops will have a
lower Hamming weight. Conversely, if the supply voltage is higher, the
propagation delay decreases, and the Hamming weight of the sample
increases. The adjustable delay stages before the carry chain
calibrate the TDC for process variation 
which ensures that the
sensor will not saturate under small voltage fluctuations that
increase or decrease the Hamming weight of the samples. {\color{black} TDC calibration
by the attacker is
required before the first time an FPGA is used for an attack or following significant
changes in device operating conditions (e.g., temperature).}
The 256-bit TDC
measurements are saved in on-chip FIFOs (256-bit word width) at
run-time and collected by the adversary after the convolution operation
is finished.

\section{Experimental Approach}
\label{sec_exp_approach}

\begin{table}[t]
\centering
\caption{\small Details of the evaluation boards used for the
experiments. The system clock generates the clock for the BNN accelerator and the TDC module.}
{\small
\begin{tabular}{| c | c | c | c |} 
 \hline
  Board Name &  Device & FPGA Family & Clk. (MHz) \\ [0.5ex] 
 \hline\hline
 ChipWhisperer &  XC7A100T & Artix 7 & 50 \\ 
 ZCU104 & XCZU7EV & Zynq UltraScale+ & 120 \\
 VCU118 & XCVU9P & Virtex UltraScale+ & 100 \\
 AWS F1 & XCVU9P & Virtex UltraScale+ & 120 \\
 \hline
\end {tabular}
}
\label{table:boards}
\end{table}

In this section, we describe the experimental platforms and
implementations used to
evaluate the efficacy of our attack. Four Xilinx FPGA-based boards,
listed in Table \ref{table:boards}, were used for experimentation. 
The first three boards in the table, locally situated in the authors' laboratories, were used
for characterization and testing. 
AWS F1 instances listed in the last row of the table were used for
cloud-based experiments.

\subsection{Experimental Platforms}

The ChipWhisperer CW305 board \cite{chipwhisperer,cw}
provides a platform for examining power side-channel
attack scenarios. The board supports low-noise off-chip voltage
measurement using an oscilloscope or a capture board via
a low-noise and high-bandwidth connection to the main FPGA 1V DC supply
pin. Voltage measurements can be obtained by an adjacent ChipWhisperer-Lite
capture board \cite{cw-lite} that contains a 10-bit analog-to-digital converter (ADC) with a 105 mega-samples per second (MS/s)
sampling rate. As described in 
Section \ref{sec_experiments}, both the capture board and an on-FPGA TDC were
used on the ChipWhisperer to obtain voltage traces. 

Xilinx ZCU104 \cite{ZCU104} and VCU118 \cite{VCU118} evaluation boards
were also 
used for evaluation, with on-FPGA TDC-based sensors used to collect voltage traces.
Off-chip FPGA supply voltage measurements were
not collected for these two boards.
The latter board contains an FPGA that is the same as the one located
on AWS F1 instances. For the AWS F1 instances, likewise, an on-FPGA TDC-based sensor was used since there is no physical access for
making off-chip supply voltage measurements.  

\subsection{Implementation on Local Boards}
\label{local-boards}

\begin{figure}
	\centering
	\includegraphics[width=0.48\textwidth]{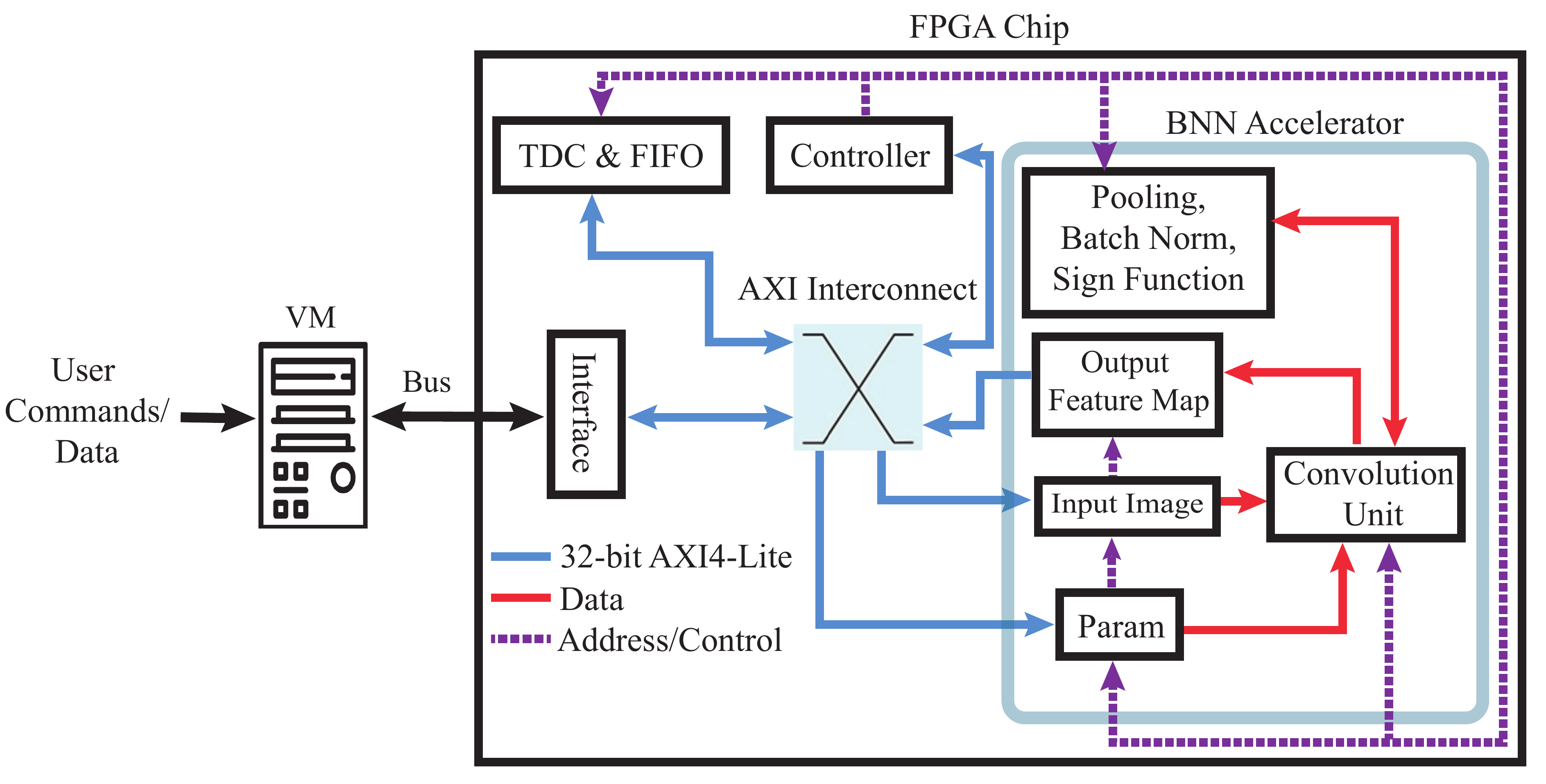}
	\caption{\small Overview of the architecture implemented on the ChipWhisper, ZCU104, VCU118 and AWS F1. 
        For AWS F1, the $Bus$ is a PCIe bus and the $Interface$ is a
Shell. For the three local boards, the $Bus$ is JTAG and the
$Interface$ is a JTAG-to-AXI converter.}
	\label{fig:overview_accelerator}
\end{figure}

The implementation of the attack architecture for the three local
boards is similar. The BNN accelerator and supporting test circuitry, as well as the TDC and FIFO used for performing the attack,  
are shown in Figure \ref{fig:overview_accelerator}. The data movement between different
components of the design takes place through an AXI4-Lite on-chip
communication protocol. Off-chip
communication (data
movement and control commands) uses a Xilinx 
JTAG-to-AXI converter module
that provides direct access to the on-chip AXI bus for the user through
a JTAG interface. 

The on-chip controller in Figure \ref{fig:overview_accelerator} sets
memory addresses and controls the operation of the convolution unit.
This controller has registers that set the parameters of the
three on-chip block memories used to store on-chip data. {\em Input Image} stores the input
feature map, {\em Output Feature Map}  stores the
result of the convolution,
and $Param$ stores the binary values of the convolution kernels
for the current layer with +1 represented by the bit value 1 and -1 by
the bit value 0. 
For each layer of the BNN, the input feature map and corresponding kernel values
are loaded into {\em Input Image} and $Param$ memories by the user, then the
convolution operation is performed, and finally the results are collected from
{\em Output Feature Map}. 

\subsection{Implementation on AWS F1}
\label{sec:awssetup}

In addition to the local boards, the attack architecture was
implemented on AWS F1 instances. 
The architecture on AWS F1 with functional modules {\em
TDC \& FIFO} and {\em BNN Accelerator} matches the 
implementations tested on the local FPGAs (Figure~\ref{fig:overview_accelerator}). 
The AWS virtual machine (VM) is able to send input images to the FPGA and read TDC
outputs from the FPGA, 
using built-in {\tt peek()} and {\tt poke()} functions.
The {\em TDC \& FIFO} modules are physically separated from the {\em BNN Accelerator} without any direct
communication.

Since AWS F1 instance FPGAs currently only support use by a single customer at a time, this setup approximates a multi-tenant scenario. Our attack does work in the presence of the $Shell$ interface circuitry and server environment that are not under user control.

\section{Attack Analysis on ChipWhisperer}
\label{sec_experiments}

In this section, we describe characterization experiments 
using the ChipWhisperer CW305 board. These experiments use both on- and
off-FPGA voltage measurements to examine voltage fluctuations
during the convolution operation as input images are processed.  The ChipWhisper is an {\em ideal} board in that its bypass capacitors have been removed and dedicated voltage measurement resources are provided.  With information gathered from the ChipWhisperer, the attack was then deployed on other, more realistic boards.

\subsection{Off-Chip Characterization of Convolution}
\label{off-chip}

\begin{figure}
	\centering
	\begin{subfigure}{.15\textwidth}
	    \includegraphics[width=1\linewidth]{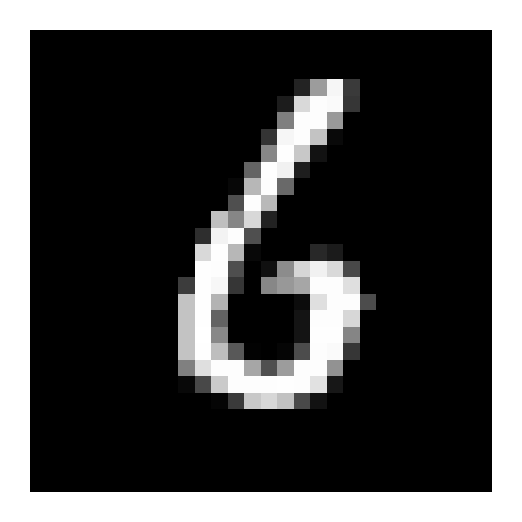}
	    \caption{}
	    \label{fig:sixes1}
	\end{subfigure}
    \begin{subfigure}{.15\textwidth}
        \includegraphics[width=1\linewidth]{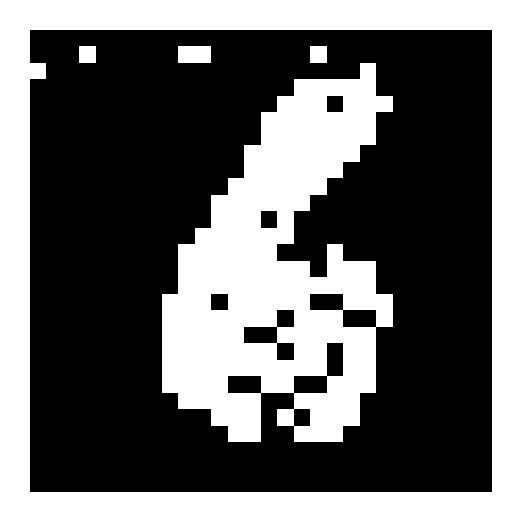}
    	\caption{}
    	\label{fig:sixes2}
    \end{subfigure}	
    \begin{subfigure}{.15\textwidth}
        \includegraphics[width=1\linewidth]{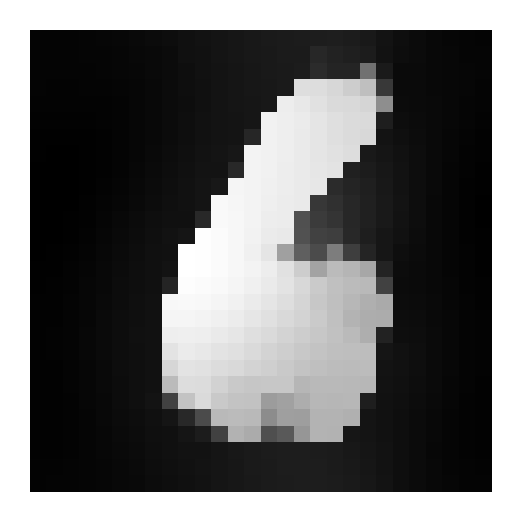}
    	\caption{}
    	\label{fig:sixes2_2}
    \end{subfigure}	
	\caption{\small (a) Input image to the convolution unit from the MNIST database, (b) recovered image with
    supply voltage traces from the ChipWhisperer board, (c) recovered image after applying a denoising algorithm.}
    \label{fig:sixes}
\end{figure}

\begin{figure}[t]
	\centering
	\includegraphics[width=0.85\linewidth]{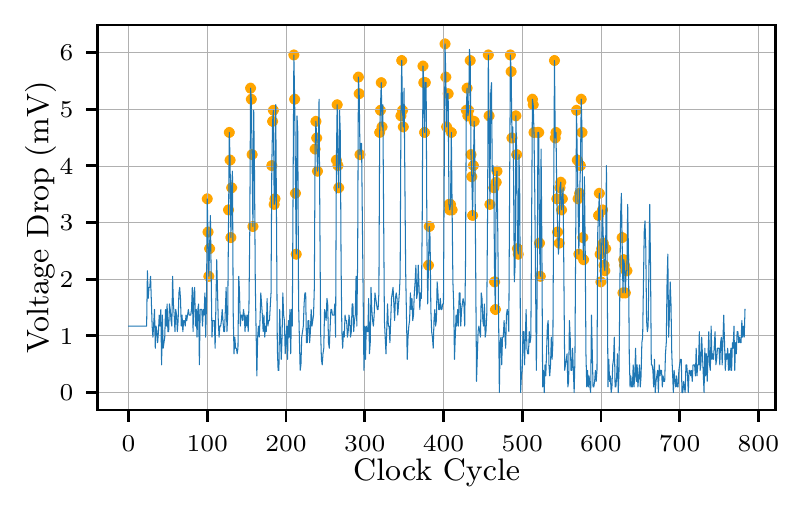}
	\caption{\small Voltage trace from the ChipWhisperer FPGA while running
the convolution unit shown in Figure \ref{fig:detailed_conv_op}. The $y$ axis
illustrates the absolute value of the measured voltage drop due to
convolution unit activity. The 125 orange circles correspond to the clock cycles
that process foreground pixels of the input image
(Figure~\ref{fig:sixes1}).} 
	\label{fig:chipwhisperer_voltage_trace}
\end{figure}

In an initial set of experiments, the ADC on the ChipWhisperer-Lite
capture board was used to sample the FPGA core supply voltage level at the
rising edge of each convolution unit clock cycle. The supply voltage level drops of
all 28$\times$28 (784) convolution operations for the first kernel
applied to the
input image, illustrated in Figure \ref{fig:sixes1}, were measured. The experiment was run 10 times
and the mean values of the voltage drops observed at the FPGA
supply input at each clock cycle versus the steady state supply voltage
were used to generate the trace shown in Figure \ref{fig:chipwhisperer_voltage_trace}.
The 125 orange circles in Figure~\ref{fig:chipwhisperer_voltage_trace}
show clock cycles during which the 125 pixels from the image foreground are
used in convolution for the first time (the clock cycle when the
foreground pixel is in location $P9$, multiplied by $K9$ in
Figure~\ref{fig:detailed_conv_op}). 
Figure~\ref{fig:chipwhisperer_voltage_trace} shows that the
clock cycles corresponding to foreground pixels have a
higher voltage drop compared to other clock cycles. These differences can be used to differentiate
between foreground and background pixels.

\begin{figure}[t]
	\centering
	\includegraphics[width=0.85\linewidth]{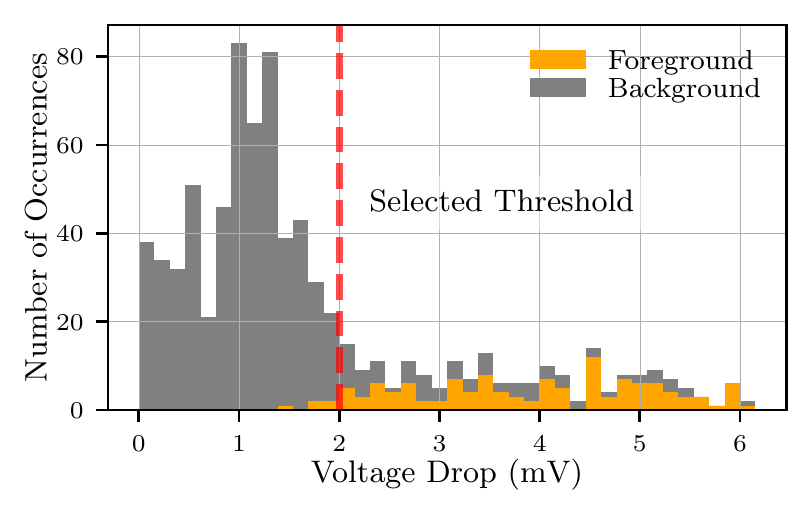}
	\caption{\small Histogram of the voltage drop due to convolution unit operation for convolution operations with the same input image and kernel. Each occurrence in the histogram represents the average of ten trials of processing the same pixel and kernel. The bars corresponding to foreground pixels are colored in orange and
those corresponding to background pixels are colored dark gray. The selected threshold (boundary) between foreground and background pixels is marked in the histogram.}
	\label{fig:chipwhisperer_voltage_hist}
\end{figure}

The voltage drops induced by the foreground pixels 
can be explained by examining Equation~\ref{eq:2}.
Each pixel of the output feature
map ($O_{x, y}$) is calculated using an image window and a kernel. 
The image (a grayscale picture of a digit with each pixel an
integer between 0 and 255) has low-valued pixels (close to 0) for background and
high-valued pixels (close to 255) for the foreground. 
For the calculation of the output feature map, the kernel values are
constant. However, the values of the processed
input image pixels in specific locations in the line buffer change between background and foreground pixels during the convolution operation. 
The dynamic power consumption (and resulting voltage
drop) of processing foreground pixels is larger than for background pixels.
Specifically, foreground pixels result in the generation of larger magnitude results for the multiply and accumulate operations when the convolution operation processes these pixels. As a result of generating these values, significant switching activity takes place in the adder tree of the convolution unit and resultant voltage drops can be observed.  

To illustrate the range of voltage changes due to the convolution of the input image, a histogram of the absolute value of voltage drop
measurements in Figure~\ref{fig:chipwhisperer_voltage_trace} is shown in
Figure~\ref{fig:chipwhisperer_voltage_hist}. The histogram contains 40 bins evenly distributed in value between 0 to 6 mV.  
The boundary between foreground and background pixels can be distinguished with a  threshold. 

Generally, the processing of background pixels leads to small voltage
drops that are clustered on the left of the histogram and the
processing of foreground pixels leads to a range of larger voltage
drops on the right of the histogram. The threshold can be identified by
locating a downward gradient in occurrence counts over multiple voltage
bins. In the ChipWhisperer, this transition took place over five bins
located just before 2 mV.

In Figure~\ref{fig:chipwhisperer_voltage_hist}, the dashed red line
shows the chosen threshold value. 
All voltage drops
created by input pixels that fall to the left of the line are
classified as background pixels, while the ones to the right are
classified as foreground pixels. To decrease noise, remove stray
pixels, and improve the quality of 
the recovered image, the Rudin-Osher-Fatemi denoising
algorithm~\cite{denoising,zhu2019} with $\tau$ equal to 0.1 and $tv\_weight$ of
40 is applied to this result to generate a recovered image. The input
image to the BNN accelerator and the two recovered images using the
threshold are shown in Figures~\ref{fig:sixes1}, \ref{fig:sixes2} and
\ref{fig:sixes2_2}, respectively. {\color{black} Unlike the
input image which has a range of grayscale pixels, the recovered images
prior to denoising are binary with 0
value for background pixels and 255 value for foreground pixels.
Following denoising, the recovered image has a range of grayscale pixels.}

\begin{figure}[t]
	\centering
	\begin{subfigure}{0.8\linewidth}
  \centering
  	\includegraphics[width=1\linewidth]{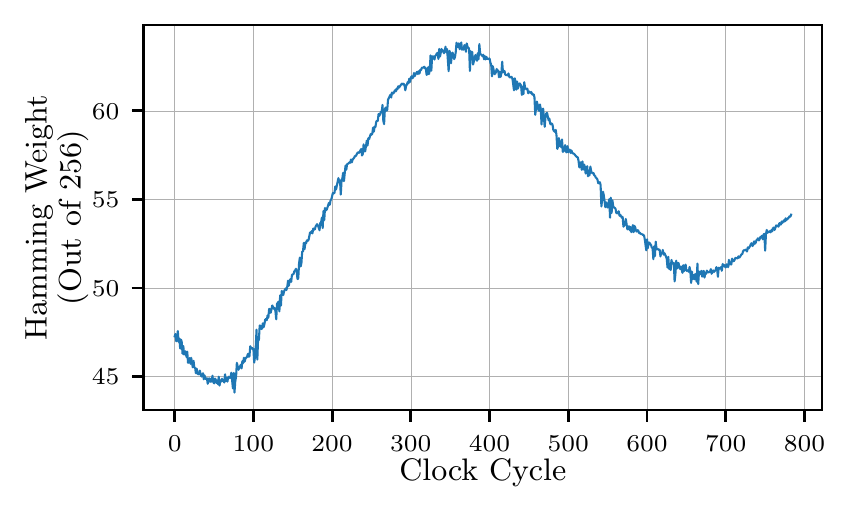}
    \caption{Average TDC Hamming weights, 100 runs.}
  \label{fig:tdc_trace_chipwhisperer_a}
\end{subfigure}%
\hfill
\begin{subfigure}{0.8\linewidth}
  \centering
  \includegraphics[width=1\linewidth]{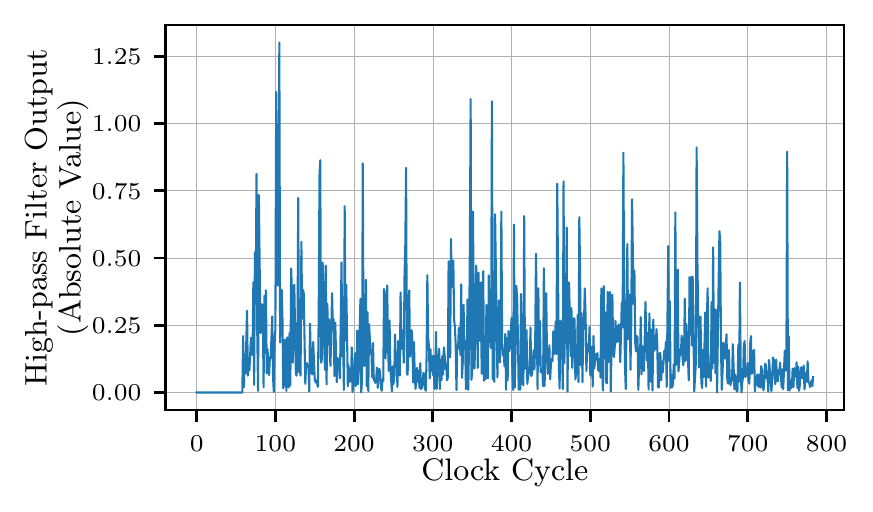}
    \caption{Recovered trace after applying a high-pass filter to the TDC Hamming weight values (absolute value).}
  \label{fig:tdc_trace_chipwhisperer_b}
\end{subfigure}
\begin{subfigure}{0.8\linewidth}
  \centering
  \includegraphics[width=1\linewidth]{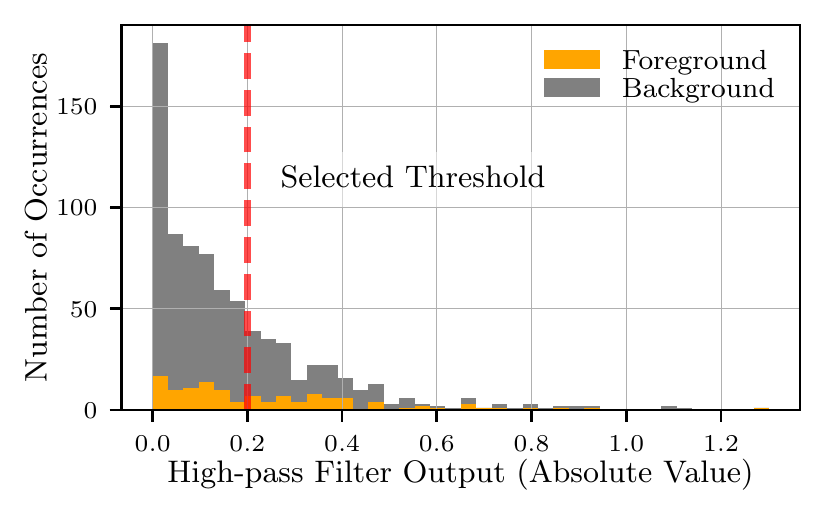}
    \caption{Histogram of the filtered TDC trace with the selected threshold shown.}
  \label{fig:tdc_trace_chipwhisperer_c}
\end{subfigure}
	\caption{\small TDC data recovered from ChipWhisperer, (a) Unfiltered trace recovered from TDC, (b) Trace after applying high-pass filter and removing low-frequency envelope (absolute value), (c) Histogram of the filtered TDC trace with the selected threshold.}
	\label{fig:tdc_trace_chipwhisperer}
\end{figure}

\subsection{TDC-Based Characterization of Convolution Operations}
\label{tdc-characterization}

The characterization of convolution unit voltage drops described in the previous subsection was performed using voltage traces obtained by the ChipWhisperer-Lite capture board. In this section, we describe
characterization experiments that use voltage measurements obtained by a TDC sensor implemented in
the ChipWhisperer FPGA. The TDC architecture was described in Section \ref{detailed-tdc}.
The 256-bit TDC carry chain for the Artix-7 FPGA on the ChipWhisperer board consists of $Carry4$ carry primitives.
The sensitivity for each TDC stage, as determined by the Xilinx Vivado 2019.1 software~\cite{vivado}, is close to 25 ps.

For each clock cycle, the flip-flop values from the TDC were saved in a 256-bit wide
FIFO, forming one voltage estimate.  This experiment was performed 100 times using the same input image and kernel. The voltage estimates at each clock cycle for the 100 traces were then averaged to minimize noise, forming a collection of 784 Hamming weights, one for each pixel. The resulting Hamming weights are shown in Figure~\ref{fig:tdc_trace_chipwhisperer_a}.

\begin{figure}[t]
	\centering
	\begin{subfigure}{0.15\textwidth}
	    \includegraphics[width=1\linewidth]{figs/input_6_trace.png}
	    \caption{}
	    \label{fig:sixes3}
	\end{subfigure}
    \begin{subfigure}{0.15\textwidth}
        \includegraphics[width=1\linewidth]{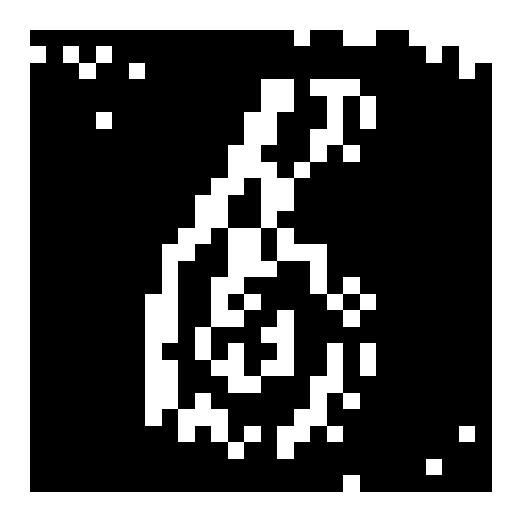}
    	\caption{}
    	\label{fig:sixes4}
    \end{subfigure}	
    \begin{subfigure}{0.15\textwidth}
        \includegraphics[width=1\linewidth]{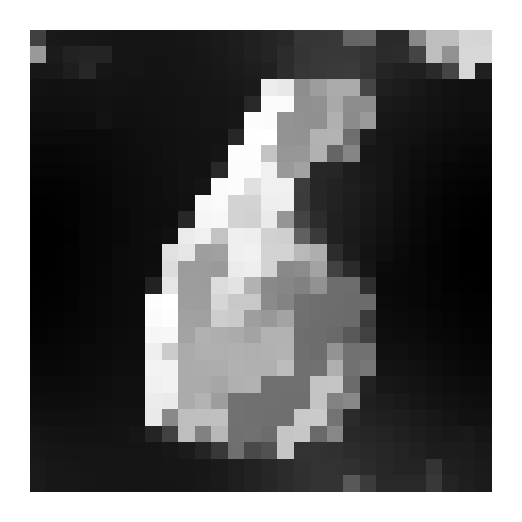}
    	\caption{}
    	\label{fig:sixes4_2}
    \end{subfigure}	
	\caption{\small Recovered image from Chipwhisperer using TDC after applying filter. (a) Input image (same as~\ref{fig:sixes1}), (b) recovered image, (c) recovered image after denoising.}
	\label{fig:tdc_chipwhisperer_recovered_6}
\end{figure}

The plot in Figure~\ref{fig:tdc_trace_chipwhisperer_a} contains a low-frequency envelope due to the lack of bypass capacitors on the ChipWhisperer that affects supply voltage behavior. 
A high-pass Butterworth digital filter was applied to the values shown in the plot to remove the envelope and retain voltage fluctuations due to convolution unit activity. For each point in the plot, the filter determines an average Hamming weight value over the previous ten clock cycles (a running average window). This value is then subtracted from the Hamming weight value at the current clock cycle, leading to the plot shown in
 Figure~\ref{fig:tdc_trace_chipwhisperer_b}. Subsequently, the image was recovered with the histogram threshold shown in Figure \ref{fig:tdc_trace_chipwhisperer_c} and Rudin-Osher-Fatemi denoising steps described earlier in Section \ref{off-chip}. 
Figure~\ref{fig:sixes4_2} shows the recovered
image obtained after applying the denoising algorithm. 

\subsection{TDC-Based Attack Summary}

To summarize, the following steps are performed to recover a reconstructed image using the on-FPGA TDC:

\begin{enumerate}
    \item Voltage estimates are collected for each input pixel during
operation of the convolution unit for the first kernel of the first
convolution layer. 
    \item Voltage estimates for each pixel are averaged across all runs
with the image to generate a single trace. The averaged estimates are represented using Hamming weights.
    \item A Butterworth high-pass filter is used to remove low-freqency power supply ripple from the averaged Hamming weights.
    \item A histogram of the resulting values is created and a threshold is used to differentiate foreground and background pixels, forming a preliminary recovered image. 
    \item A Rudin-Osher-Fatemi denoising algorithm is used to improve the quality of the recovered image.
    
\end{enumerate}

\begin{figure}[t]
\centering

\begin{subfigure}[b]{0.49\textwidth}
\centering
   \includegraphics[trim={0 0.8cm 0 0}, clip, width=\linewidth]{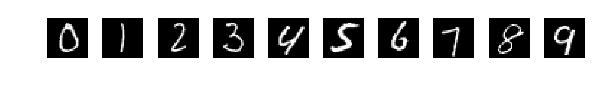}
   \caption{Input images.}
   \label{fig:Ng1} 
\end{subfigure}
\\
\begin{subfigure}[b]{0.49\textwidth}
\centering
   \includegraphics[trim={0 0.8cm 0 0}, clip, width=\linewidth]{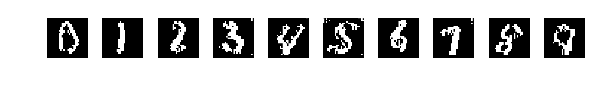}
   \caption{Recovered images from ZCU104 before denoising.}
   \label{fig:Ng4} 
\end{subfigure}
\begin{subfigure}[b]{0.49\textwidth}
\centering
   \includegraphics[trim={0 0.8cm 0 0}, clip, width=\linewidth]{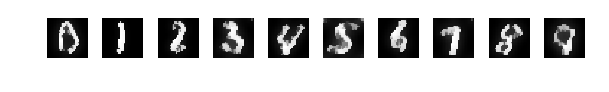}
   \caption{Recovered images from ZCU104 after denoising.}
   \label{fig:Ng5}
\end{subfigure}
\begin{subfigure}[b]{0.49\textwidth}
\centering
   \includegraphics[trim={0 0.8cm 0 0}, clip, width=\linewidth]{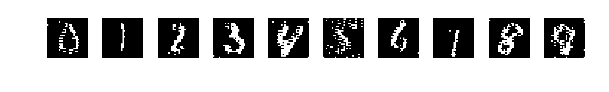}
   \caption{Recovered images from VCU118 before denoising.}
   \label{fig:Ng6} 
\end{subfigure}
\begin{subfigure}[b]{0.49\textwidth}
\centering
   \includegraphics[trim={0 0.8cm 0 0}, clip, width=\linewidth]{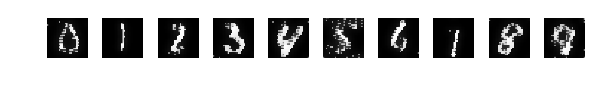}
   \caption{Recovered images from VCU118 after denoising.}
   \label{fig:Ng7} 
\end{subfigure}
\begin{subfigure}[b]{0.49\textwidth}
\centering
   \includegraphics[trim={0 0.8cm 0 0}, clip, width=\linewidth]{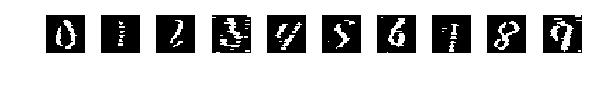}
   \caption{Recovered images from AWS F1 before denoising.}
   \label{fig:Ng8} 
\end{subfigure}
\begin{subfigure}[b]{0.49\textwidth}
\centering
   \includegraphics[trim={0 0.8cm 0 0}, clip, width=\linewidth]{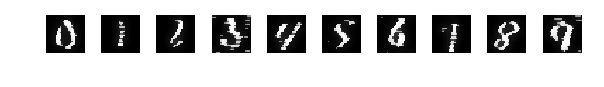}
   \caption{Recovered images from AWS F1 after denoising.}
   \label{fig:Ng9} 
\end{subfigure}
\caption{\small Input images and recovered images before and after denoising from all boards. The images were recovered using only TDC measurements.}
\label{fig:recovered_all_figures}
\end{figure}

\section {Image Extraction Using the Attack}
\label{image_extract}

After initial experimentation with the ChipWhisperer CW305, our attack
was applied to the two local boards and AWS F1 instances described in
Section \ref{sec_exp_approach} to see how well the attack can perform
on commercial off-the-shelf boards that were not designed to study side
channel attacks. The hardware for these platforms was not modified for
our experimentation. The experimental setup for these platforms
including the BNN accelerator is shown in Figure
\ref{fig:overview_accelerator}. The clock speeds
of the BNN accelerators in the FPGAs are listed in Table
\ref{table:boards}. Our experiments consider the quality of the
recovered images, the proximity of the TDC to the convolution unit on
the FPGA chip, and the number of times each input image is used to
create a recognizable recovered image (e.g., number of runs).

\subsection{Image Recovery with Local Boards}
\label{adjacent}

\begin{figure}[t]
	\centering
	\begin{subfigure}{0.20\textwidth}
    	\includegraphics[width=1\linewidth]{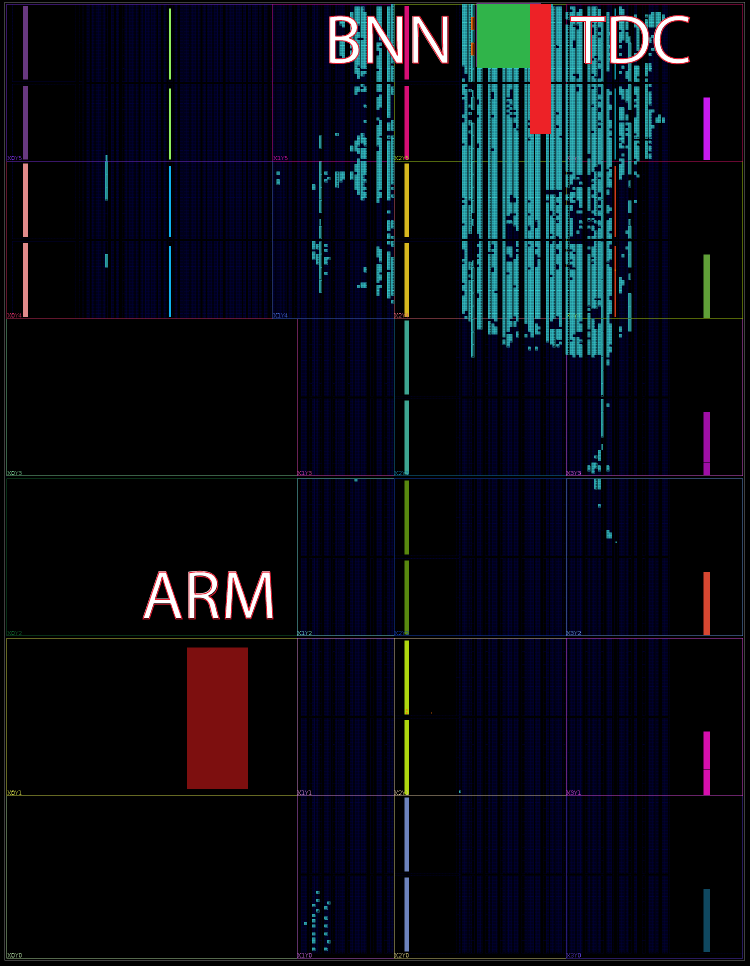}    
        \caption{Adjacent Placement}
        \label{fig:floorplan1}
    \end{subfigure}
    \begin{subfigure}{0.20\textwidth}
    	\includegraphics[width=1\linewidth]{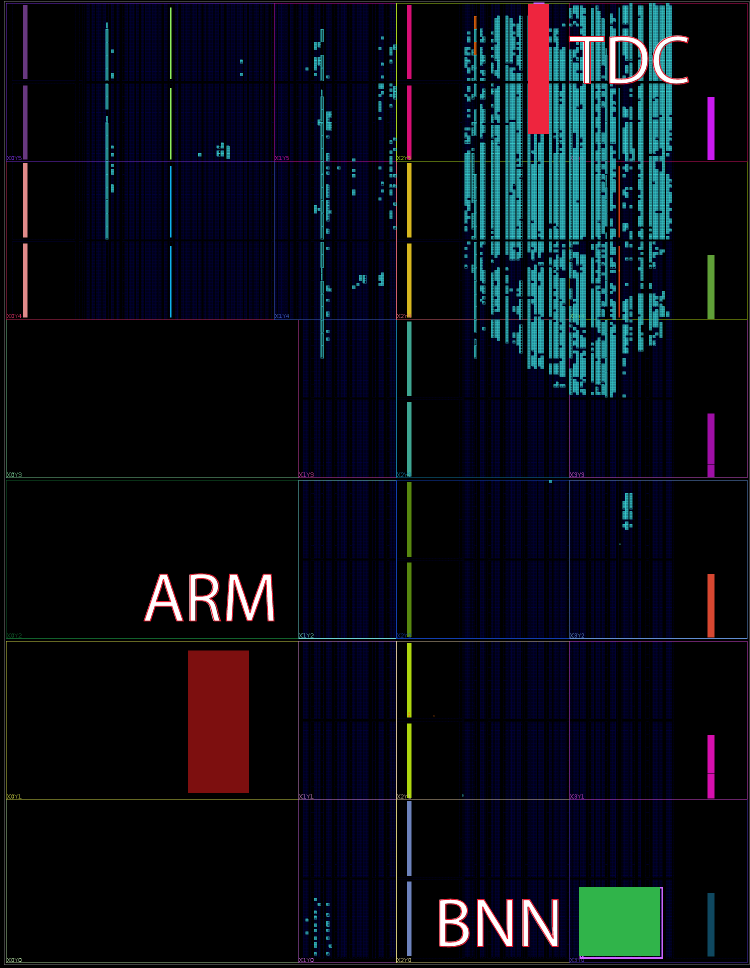}
        \caption{Cross-Die Placement} 
        \label{fig:floorplan2}
	\end{subfigure}
	        \caption{\small Floorplan of the ZCU104 UltraScale+ FPGA
for adjacent and cross-die placement. Green rectangle: BNN accelerator.
Red rectangle: TDC sensor. {\color{black} Brown rectangle: ARM processor.}}
	    \label{fig:floorplan}
\end{figure}

Recovered images, both before and after denoising, for the ZCU104 and VCU118 boards using TDC measurements are shown in Figure \ref{fig:recovered_all_figures}. For these experiments, the TDC was placed adjacent to the BNN accelerator in the FPGA
fabric (in the next row of logic blocks) to increase the accuracy of the voltage estimates. For example, the relative positions of the BNN accelerator and TDC in the ZCU104's UltraScale+ FPGA for these experiments are shown in Figure \ref{fig:floorplan1}. The images were recovered after applying the steps outlined in Section \ref{tdc-characterization}. For the ZCU104 and VCU118, the same input image and kernel were used 3,000 times.

The TDC's ability to detect the small voltage drops caused by the convolution unit as it processes the input image is critical to image recovery. 
To study the importance of TDC location on the FPGA die relative to the location of the BNN accelerator, the BNN was moved to a location on the opposite side of the die, as shown in Figure \ref{fig:floorplan2} for the ZCU104's UltraScale+ FPGA. The experiments from Section \ref{adjacent} were rerun for the digital image shown in 
Figure~\ref{fig:sixes1}. 

 To compare the quality of the recovered images with cross-die
placement of the TDC versus the results from adjacent placement for
the selected digit, the normalized cross-correlation ($CCR\_norm$), derived from cross-correlation ($CCR$),
between the recovered images and the input image for both adjacent and
cross-die TDC placements were calculated using Equations \ref{ccr} and
\ref{ccr_norm}. Here,~$\bar{{A}}$ and $\bar{{B}}$ represent the mean
pixel values of the images. 
The $CCR\_norm$ value provides a quantitative metric
for comparing the similarity of the input image and a recovered image.  
\begin{equation}
\label{ccr}
CCR = \sum_{(i, j) \in N^{28 \times 28}}\big[(A[i, j]-\bar{{A}})\times (B[i, j]-\bar{{B}})\big] 
\end{equation}

\begin{equation}
\label{ccr_norm}
CCR\_norm = \frac{CCR}{\sqrt{\sum \big(A[i, j]-\bar{A}\big)^{2} \times \sum \big(B[i, j]-\bar{B}\big)^{2}}}
\end{equation}

\begin{figure}[htbp!]
	\centering
	\begin{subfigure}{0.10\textwidth}
	    \centering
	    \captionsetup{justification=centering}
	    \includegraphics[width=0.8\linewidth]{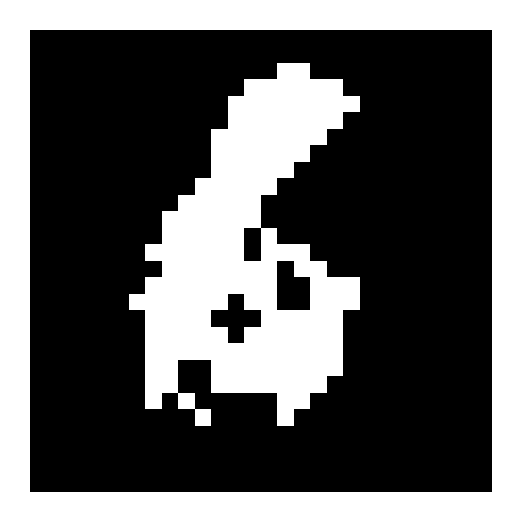}
	    \caption{\small ZCU104,\\cross-die,\\w/o denoising}
	    \label{fig:isolate_1}
	\end{subfigure}
    \begin{subfigure}{0.10\textwidth}
    	\centering
	\captionsetup{justification=centering}
        \includegraphics[width=0.8\linewidth]{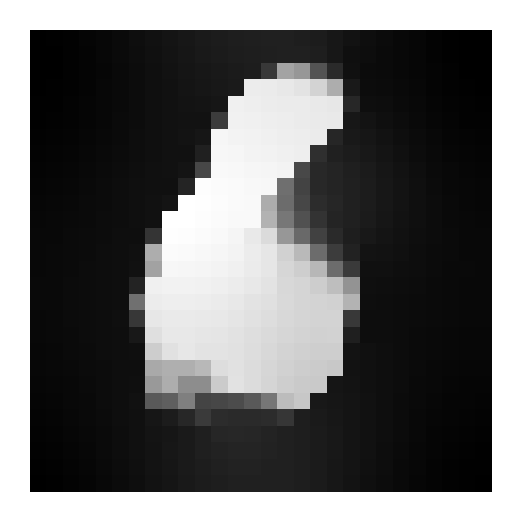}
    	\caption{\small ZCU104,\\cross-die,\\denoising}
    	\label{fig:isolate_2}
    \end{subfigure}	
	\begin{subfigure}{0.10\textwidth}
	    \centering
	    \captionsetup{justification=centering}
	    \includegraphics[width=0.8\linewidth]{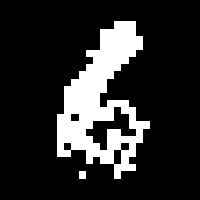}
	    \caption{\small VCU118,\\cross-die,\\w/o denoising}
	    \label{fig:isolate_3}
	\end{subfigure}
    \begin{subfigure}{0.10\textwidth}
        \centering
        \captionsetup{justification=centering}
        \includegraphics[width=0.8\linewidth]{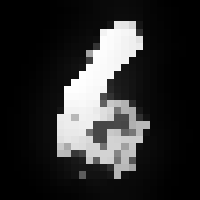}
    	\caption{\small VCU118,\\cross-die,\\denoising}
    	\label{fig:isolate_4}
    \end{subfigure}\\	

	\centering
	
    \begin{subfigure}{0.10\textwidth}
        \centering
        \captionsetup{justification=centering}
        \includegraphics[width=0.8\linewidth]{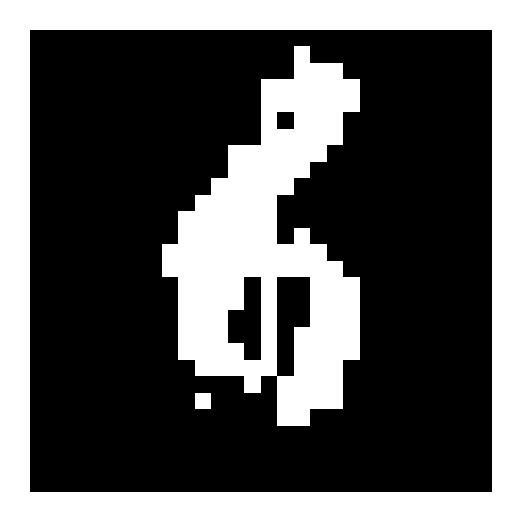}
    	\caption{\small ZCU104,\\adjacent,\\w/o denoising}
    	\label{fig:close_1}
    \end{subfigure}	
    \begin{subfigure}{0.10\textwidth}
        \centering
        \captionsetup{justification=centering}
        \includegraphics[width=0.8\linewidth]{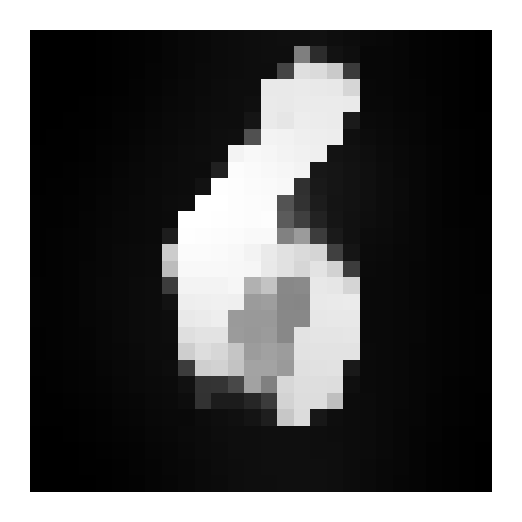}
    	\caption{\small ZCU104,\\adjacent,\\denoising}
    	\label{fig:close_2}
    \end{subfigure}
    \begin{subfigure}{0.10\textwidth}
        \centering
        \captionsetup{justification=centering}
        \includegraphics[width=0.8\linewidth]{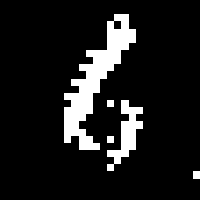}
    	\caption{\small VCU118,\\adjacent,\\w/o denoising}
    	\label{fig:close_3}
    \end{subfigure}	
    \begin{subfigure}{0.10\textwidth}
        \centering
        \captionsetup{justification=centering}
        \includegraphics[width=0.8\linewidth]{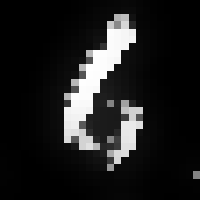}
    	\caption{\small VCU118,\\adjacent,\\denoising}
    	\label{fig:close_4}
    \end{subfigure}
	\caption{\small {\color{black} Recovered images with adjacent and cross-die placement
for 3,000 runs. The input image is shown in~\ref{fig:sixes1}.}}
	\label{fig:close_vs_isolate_six}
\end{figure}

\noindent Recovered images both before and after denoising were considered.
Table~\ref{table:3} shows the normalized cross-correlations of the
recovered images on the target boards. Figure~\ref{fig:close_vs_isolate_six} shows the recovered images for different placement strategies, both before and after denoising, for the two boards.

\begin{table}[t]
\centering
\caption{\small Normalized cross-correlation between original
and recovered images before and after denoising under
adjacent and cross-die TDC placement. 
The ZCU104 and AWS F1 FPGA floorplans are shown in Figures
\ref{fig:floorplan} and \ref{fig:awsfloorplan}. 
}
{\small
\begin{tabular}{| c | c | c | c | c |} 
 \hline
  Board &   \multicolumn{2}{c|}{Adjacent Placement }& \multicolumn{2}{c|}{Cross-die Placement}  \\ 
 \hline\hline
  & w/o denoise & w/ denoise & w/o denoise & w/ denoise \\
  \hline
 ZCU104 & 0.745& 0.791 & 0.594 & 0.655 \\
 VCU118 & 0.678& 0.738 & 0.646 & 0.697 \\
 AWS F1 & 0.671& 0.716 & 0.426 & 0.547 \\ 
 \hline
\end {tabular}
}
\label{table:3}
\end{table}

This experiment shows that cross-die placement leads to the recovery of a lower-quality image compared to adjacent placement, which was predictable. However,
the recovered image is still recognizable and the attack can
be performed even if the BNN accelerator and TDC are not in close proximity.

\begin{figure}[t]
	\centering
	\includegraphics[width=0.9\linewidth]{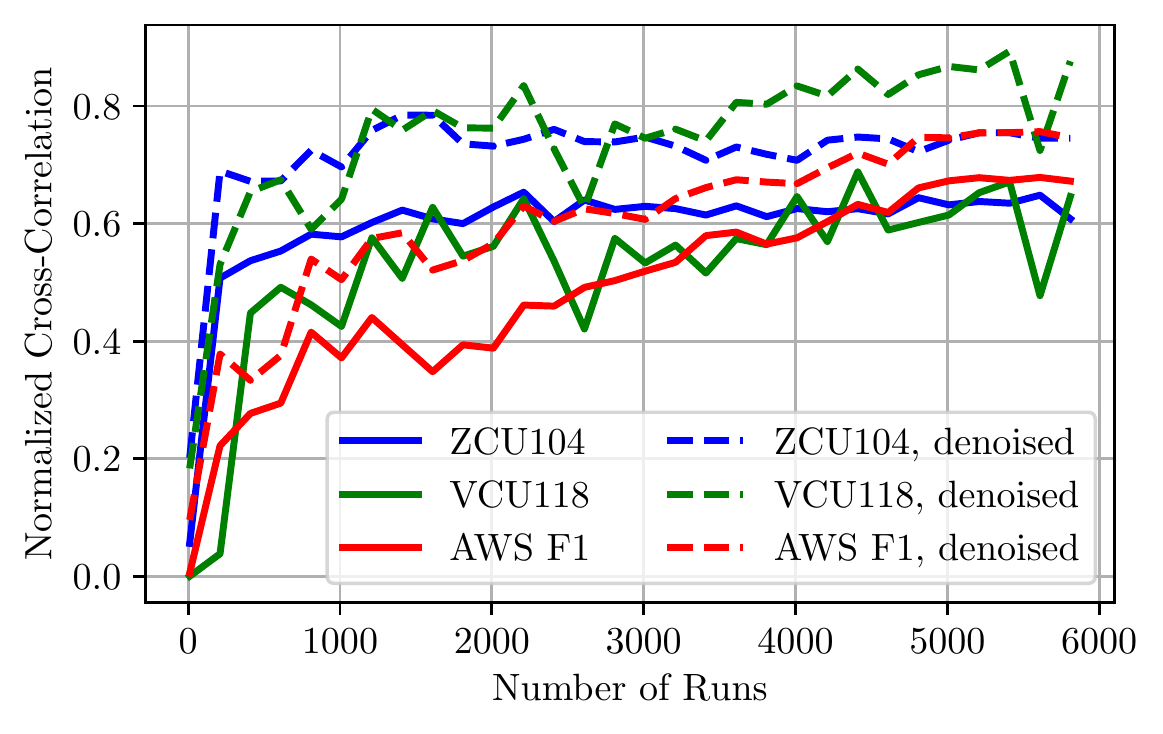}
	\caption{\small Normalized cross correlation of the recovered image versus number of runs
for all boards, before and after denoising. Input image shown in
Figure~\ref{fig:sixes1}.} 

	\label{fig:ccr_vs_run}
\end{figure}

\begin{figure}[t]
	\centering
	\begin{subfigure}[b]{0.23\textwidth}
	    \centering
	    \includegraphics[width=\linewidth]{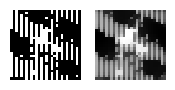}
	    \caption{\small 100 runs, 0.186 $\rightarrow$ 0.365}
	    \label{fig:six_runs1}
	\end{subfigure}
	\begin{subfigure}[b]{0.23\textwidth}
	    \centering
	    \includegraphics[width=\linewidth]{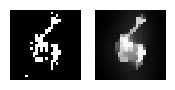}
	    \caption{\small 200 runs, 0.592 $\rightarrow$ 0.739}
	    \label{fig:six_runs2}
	\end{subfigure}
	\begin{subfigure}[b]{0.23\textwidth}
	    \centering
	    \includegraphics[width=\linewidth]{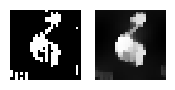}
	    \caption{\small 500 runs, 0.611 $\rightarrow$ 0.692}
	    \label{fig:six_runs3}
	\end{subfigure}
	\begin{subfigure}[b]{0.23\textwidth}
	    \centering
	    \includegraphics[width=\linewidth]{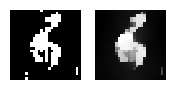}
	    \caption{\small 1,000 run, 0.653 $\rightarrow$ 0.741}
	    \label{fig:six_runs4}
	\end{subfigure}
	\begin{subfigure}[b]{0.23\textwidth}
	    \centering
	    \includegraphics[width=\linewidth]{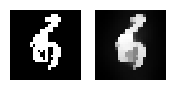}
	    \caption{\small 3,000 runs, 0.744 $\rightarrow$ 0.785}
	    \label{fig:six_runs5}
	\end{subfigure}
	\begin{subfigure}[b]{0.23\textwidth}
	    \centering
	    \includegraphics[width=\linewidth]{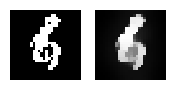}
	    \caption{\small 6,000 runs, 0.735 $\rightarrow$ 0.779}
	    \label{fig:six_runs6}
	\end{subfigure}
	\caption{\small Recovered images for the ZCU104 board for different numbers of runs before (left figure) and after (right figure) denoising. Normalized cross-correlation with the original image is included in each caption ($CCR\_norm$ without denoising $\rightarrow$ $CCR\_norm$ with denoising). Input image shown in Figure~\ref{fig:sixes1}.
	}
	\label{fig:six_runs}
\end{figure}

To obtain recognizable reconstructed images, the same input image is processed by the same kernel numerous times. To evaluate the effect of number of runs on image quality, we again used the image shown in Figure \ref{fig:sixes1}.
For both local FPGA boards, the normalized cross-correlation (Equation~\ref{ccr_norm}) of the recovered image and the original image versus the number of times the input image was processed by the first kernel was calculated. Results from these experiments are shown in Figure~\ref{fig:ccr_vs_run}. Denoising the recovered images clearly improves the image quality.  Figure~\ref{fig:six_runs} shows recovered images for an increasing number of runs, before and after denoising. This figure clearly shows that after about 200 runs, the recovered image is recognizable.

\begin{figure}[t]
	\centering
	\includegraphics[width=0.9\linewidth]{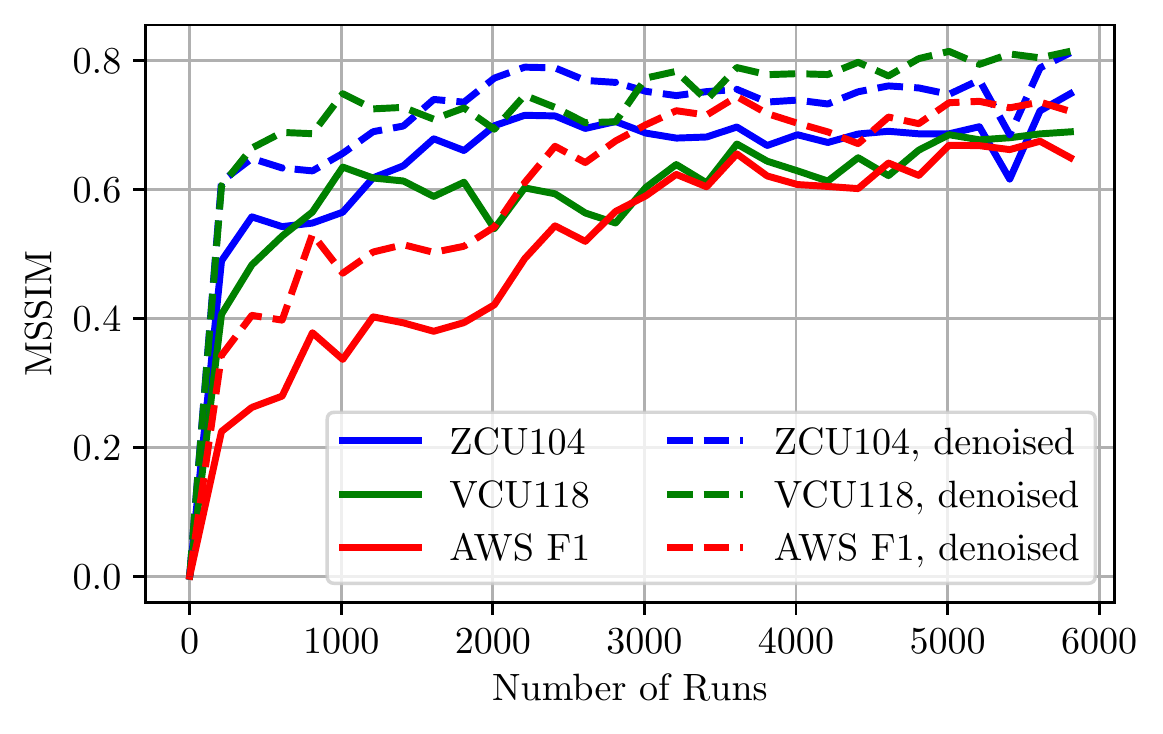}
	\caption{{\color{black} \small Mean structural similarity (MSSIM) of the
recovered image versus number of runs for all boards, before and after denoising.
Input image shown in Figure~\ref{fig:sixes1}.}}

	\label{fig:mssim}
\end{figure}

{\color{black}
\subsection{Analysis of Image Mean Structural Similarity}
To further contrast the perceptual similarity of recovered and
original input images, the mean structural similarity index
(MSSIM~\cite{SSIM}) was calculated. 
The MSSIM of two images is determined by taking the mean of the
structural similarity index values between fixed-size windows of the
two images rather than comparing individual pixels. Structural
similarity index provides a quantitative comparison between the two
image windows. 
MSSIM calculations for two images generate a value between -1 and 1,
with values close to 1 indicating a close match and values close to -1
indicating a complete mismatch. A sliding window size of 11 pixels was
chosen for calculating MSSIM values \cite{SSIM}.
The mean structural similarity index between the input
image in Figure~\ref{fig:sixes1} and the recovered image for different numbers of
runs is shown in Figure~\ref{fig:mssim}. The plots in the figure closely follow
the normalized cross correlation trends shown in Figure~\ref{fig:ccr_vs_run} as the MSSIM
index increases when the number of runs
increases.  
}

\subsection{Effect of Voltage Stressing Circuits on Local Board Image Recovery}

\begin{figure}
    \centering
    \begin{tabular}{cc}
        \begin{subfigure}[b]{0.33\linewidth}
        \centering
           \includegraphics[width=0.88\linewidth]{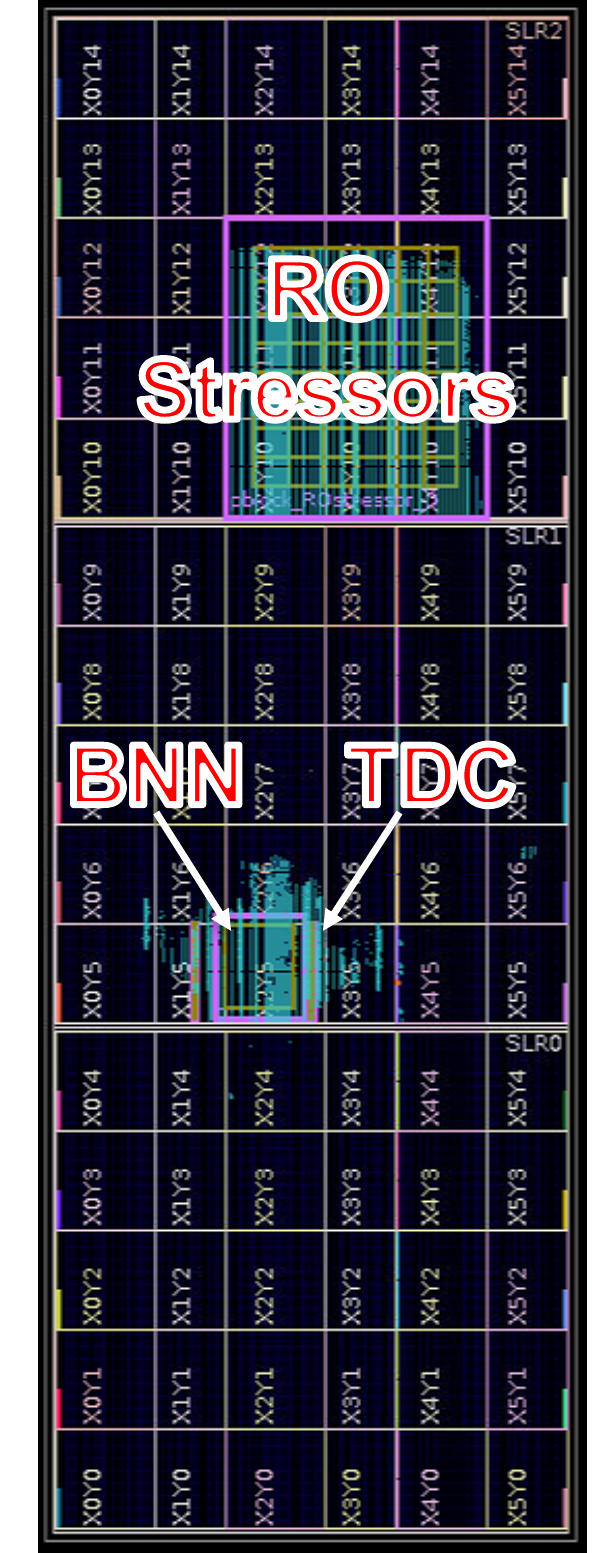}
           \caption{Floorplan of experiment with RO stressors}
           \label{fig:rostressor_floorplan} 
        \end{subfigure}
        &      
        \adjustbox{valign=b}{
            \adjustbox{valign=t}{\begin{tabular}{@{}c@{}}
  
  \begin{subfigure}[t]{0.4\linewidth}
                \centering
                   \includegraphics[width=\linewidth]{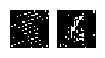}
                   \caption{100 runs}
                   \label{fig:rostressor_ro3} 
                \end{subfigure} \\
                
                \begin{subfigure}[t]{0.4\linewidth}
                \centering
                   \includegraphics[width=\linewidth]{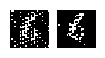}
                   \caption{500 runs}
                   \label{fig:rostressor_ro2} 
                \end{subfigure}\\
   
   \begin{subfigure}[b]{0.4\linewidth}
                \centering
                   \includegraphics[width=\linewidth]{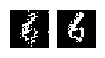}
                   \caption{1,000 runs}
                   \label{fig:rostressor_ro1} 
                \end{subfigure}
                
           \end{tabular}}}
    \end{tabular}
\caption{\small The effects of instantiated stressor circuits on local VCU118. In (b) - (d), the images recovered without and with 50,000 stressors are shown on the left and right, respectively.}
\label{fig:rostressor}
\end{figure}

It has previously been shown that an attacker's ability to detect small
on-FPGA voltage changes is enhanced if significant steady-state power
is simultaneously drawn from the device \cite{giechaskiel2020capsule}.
In addition to the TDC and associated control circuitry, an attacker
may instantiate circuits that deliberately consume significant power in
an effort to stress the power distribution network of the supply
voltage. A common voltage stressor circuit is a ring oscillator (RO), a
combinational loop that contains an odd number of inverters. This type
of stressor can be efficiently implemented in an FPGA using one logic
element.

In an experiment with the VCU118, RO-based voltage stressors were added to the UltraScale+ FPGA and enabled during the extraction of voltage estimates from the convolution of the input image and first kernel. 
As shown in the floorplan in Figure~\ref{fig:rostressor_floorplan},
the TDC and BNN accelerator were located in adjacent columns on the device and
the stressors were located in a different region of the device to reduce their effect on on-die temperature. 
Fifty groups of RO stressors were used, each with 1,000 ROs, for a total of 50,000 (shown in Figure \ref{fig:rostressor_floorplan}). This stressor count was found to be sufficient to impact the appearance of the recovered images.

To examine effects of using stressors, the image shown in Figure \ref{fig:sixes1} was input into the FPGA for separate experiments in which the stressors were activated or not activated. The recovered images for an increasing number of runs during the experiments are shown at the right in Figure \ref{fig:rostressor}. The images indicate the visual improvement as a result of stressor deployment.

\subsection{Image Reconstruction on AWS F1}


\begin{figure}[t]
    \centering
        \begin{subfigure}[b]{0.8\linewidth}
           \centering
           \includegraphics[width=\linewidth]{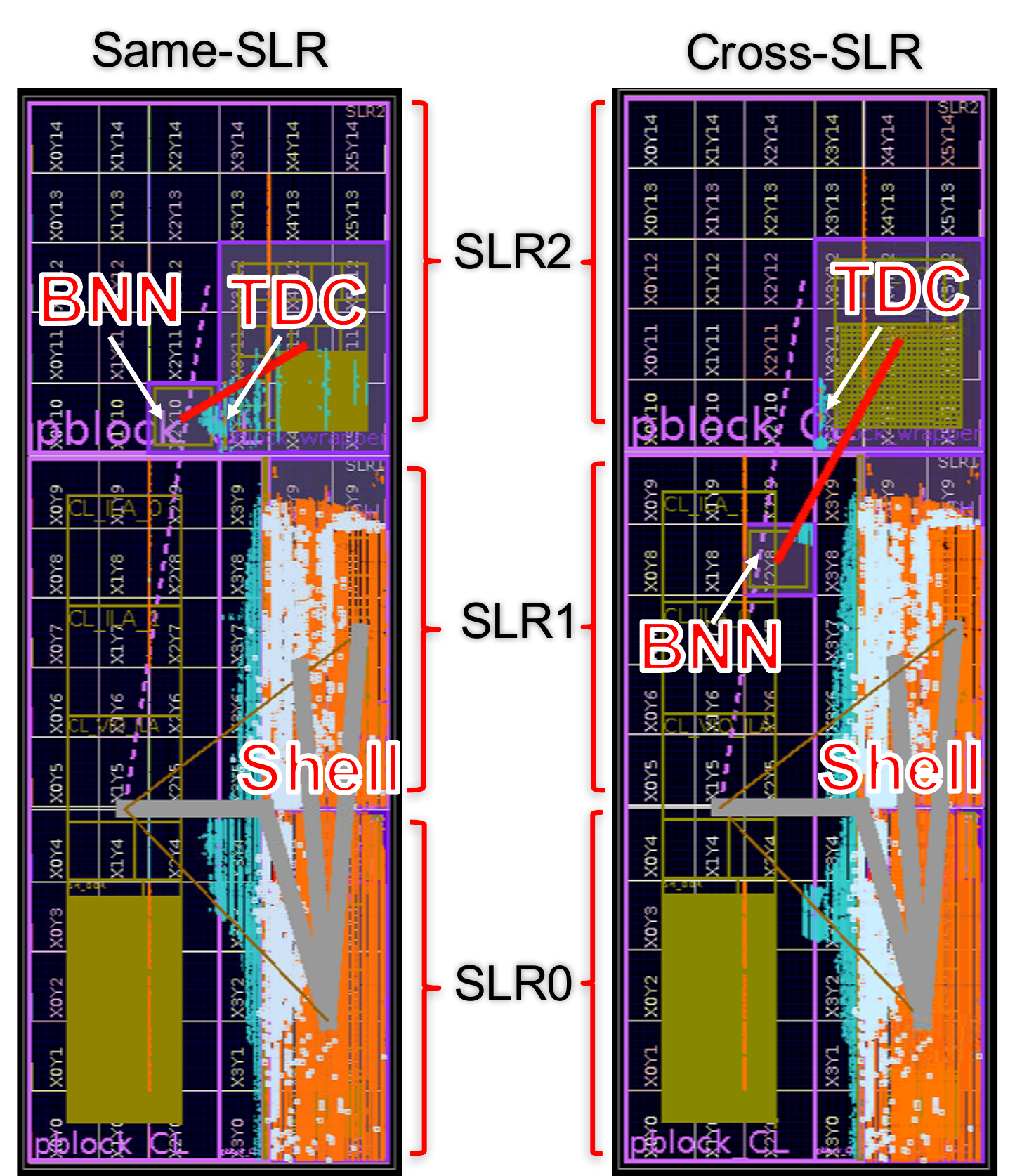}
           \captionsetup{justification=centering,margin=0.1cm}
           \caption{\small Floorplan of same-SLR and cross-SLR experiments on AWS F1}
           \label{fig:awsfloorplan}
        \end{subfigure}
        
        \begin{tabular}{cc}
            \begin{subfigure}[]{0.23\textwidth}
               \centering
               \includegraphics[width=0.45\linewidth]{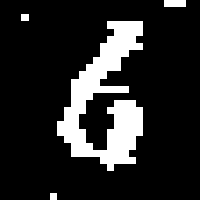}
               \includegraphics[width=0.45\linewidth]{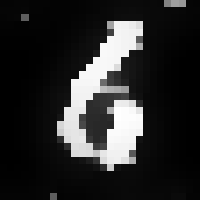}
               \caption{\small Same-SLR, w/o denoising (left) and w/ denoising (right)}
               \label{fig:aws_adjacent_image} 
            \end{subfigure}
       
             \begin{subfigure}[]{0.23\textwidth}
               \centering
               \includegraphics[width=0.45\linewidth]{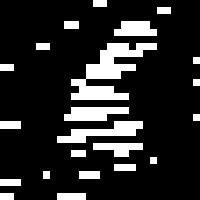}
               \includegraphics[width=0.45\linewidth]{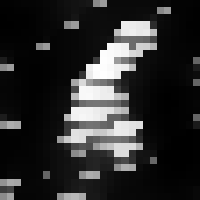}
               \caption{\small Cross-SLR, w/o denoising (left) and w/ denoising (right)}
               \label{fig:aws_cross_image} 
            \end{subfigure}
        \end{tabular}
\caption{\small Floorplan and recovered images of same-SLR and cross-SLR experiments on AWS F1: (a) The {\em Shell} logic occupies the right part of SLR0 and SLR1. Left: The TDC and BNN accelerator are both on SLR2; 
Right: The TDC is on SLR2, and the accelerator BNN is on SLR1. 
(b) The recovered image of same-SLR experiment on AWS F1 for 6,000 runs.
(c) The recovered image of cross-SLR experiment on AWS F1 for 6,000 runs.
}
\label{fig:aws_floorplan_digits}
\end{figure}

\begin{figure}[t]
\centering
\begin{subfigure}[t]{0.8\linewidth}
\centering
   \includegraphics[width=\linewidth]{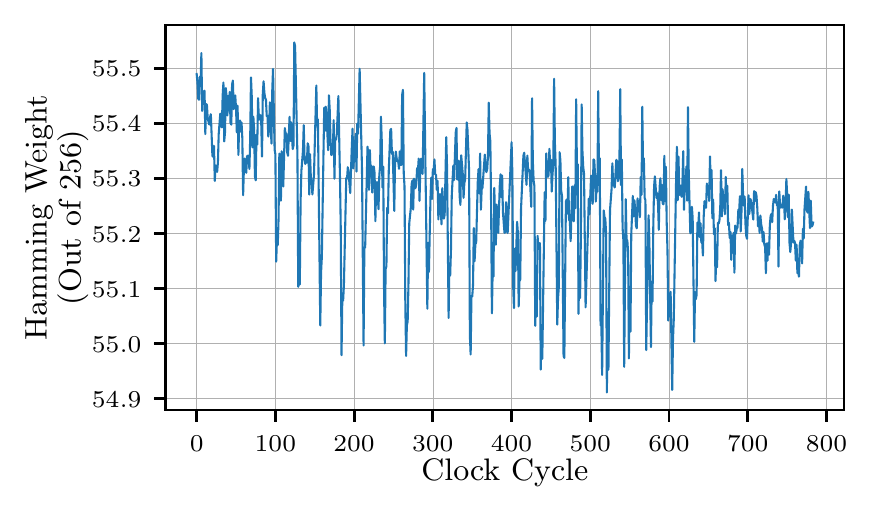}
   \caption{\small Average TDC Hamming weights, 6,000 runs.}
   \label{fig:awshpf_raw} 
\end{subfigure}
\begin{subfigure}[t]{0.8\linewidth}
\centering
   \includegraphics[width=\linewidth]{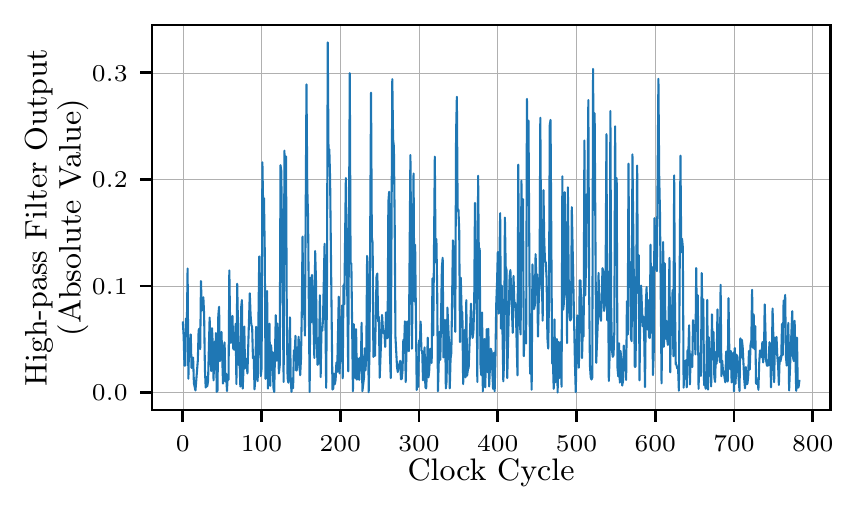}
   \caption{\small Recovered trace after applying a high-pass filter to the TDC
Hamming weight values (absolute value).}
   \label{fig:awshpf_filter} 
\end{subfigure}
\begin{subfigure}[t]{0.8\linewidth}
\centering
   \includegraphics[width=\linewidth]{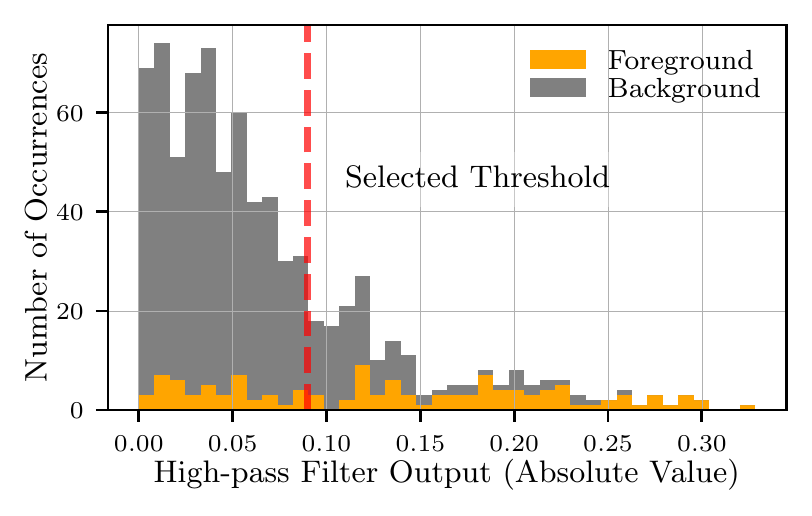}
   \caption{\small Histogram of the filtered TDC trace with the selected threshold shown.}
   \label{fig:awshpf_filter_hist} 
\end{subfigure}
\caption{\small 
Average of TDC voltage drop traces averaged over 6,000 runs on AWS F1 for same-SLR experiment. (a) Unfiltered
trace recovered from the TDC; (b) Trace after applying high pass filter (absolute value); (c) Histogram of the filtered TDC trace and the selected threshold.
}
\label{fig:awshpf}
\end{figure}

To show that our attack could be deployed on existing cloud FPGAs if multi-tenancy was allowed, our attack infrastructure was migrated to and tested on AWS F1 instances. The experimental setup for the attack was described in Section \ref{sec:awssetup}.
The UltraScale+ FPGA used in AWS F1 contains three {\em super logic regions} (SLRs) (Figure~\ref{fig:awsfloorplan}).
Each SLR is a separate die containing logic and memory resources. As shown in Figure~\ref{fig:awsfloorplan}, the {\em Shell} interface is located in the right-hand area of SLR0 and SLR1. 
Since the {\em Shell} has significant power consumption, it can influence the accuracy of TDC measurements. 
To assess these effects, experiments were performed with the BNN accelerator and TDC on SLR2 (same-SLR) and on separate SLRs (cross-SLR).
In the same-SLR experiment (Figure \ref{fig:awsfloorplan} left), the TDC and the BNN are placed next to each other.
Figure~\ref{fig:awsfloorplan} (right) depicts the cross-SLR experiment, in which the BNN accelerator is on SLR1 and the TDC is on SLR2.

Figure~\ref{fig:awshpf} shows the averaged Hamming weights obtained for the same-SLR case using the digit image from Figure \ref{fig:sixes1} for 6,000 runs. 
As shown in Figure~\ref{fig:awshpf_raw}, the averaged values collected by the TDC are influenced by environmental noise, a decreasing voltage envelope. After high-pass filtering, identifiable peaks, indicating foreground pixels, can be identified,
as shown in Figure~\ref{fig:awshpf_filter}. The histogram and selected threshold used to extract the recovered image are shown in Figure \ref{fig:awshpf_filter_hist}. 

The recovered images of the same-SLR experiment are shown in Figures~\ref{fig:Ng8} and \ref{fig:Ng9}. 
The normalized cross-correlations between the input and recovered images for the same-SLR case before and after denoising are 0.671 and 0.716, respectively. 
As shown in Figure~\ref{fig:ccr_vs_run}, the normalized cross-correlations for denoised images for the same-SLR experiment on AWS F1 increase as the number of runs are increased.

Similar to the local board experiments, the positioning of the TDC at a distant location from the BNN accelerator results in a reduction in recovered image quality. 
The experiment described in the previous paragraph was rerun on AWS F1 for the cross-SLR case.
The recovered images of digit 6 are shown in Figure~\ref{fig:aws_cross_image},
and the averages of normalized cross-correlation are listed in Table~\ref{table:3}.
The results indicate that the normalized cross-correlation of the denoised image for the same-SLR experiment (0.716) is superior to the value for the cross-SLR experiment (0.547). The presence of the $Shell$ in the same SLR as the BNN accelerator for the cross-die experiment influences the quality of the recovered image to a modest extent.

{\color{black}
\subsection{Limitations}

Although our image reconstruction attack has been shown to be
effective on multiple FPGA-based boards, it does have
limitations. All presented results thus far were generated using
the MNIST handwritten digit database which includes images with background pixels of 0 and
foreground pixels with values up to 255. Additionally, our previously
reconstructed images have used multiple repetitions (runs) with the
exact same image. In this subsection, we
examine the performance of the attack on ZCU104 and
VCU118 boards if these constraints are relaxed. 

As described in Section \ref{sec_experiments}, the use of pixel values of 0 for the
background minimizes adder tree activity in Figure~\ref{fig:detailed_conv_op}, leading to a significant
difference in voltage drops caused by foreground and background
pixels. For experimentation, four new versions of the image
shown in Figure \ref{fig:sixes1} were created with all background pixels converted to 1, 10, 30,
and 50, respectively. Reconstructed images were generated for each
modified input image after 6,000 runs each. Table \ref{table:4} indicates that
deviation from a zero-valued background does indeed reduce normalized
cross-correlation in all cases, although, as shown in Figure \ref{fig:example_background}, the
reconstructed images are still recognizable after denoising. 

\begin{table}[t]
\caption{{\color{black} Normalized cross-correlation of the recovered image and
the original image shown in Figure~\ref{fig:sixes1} after replacing the
background pixels with a non-zero constant value prior to BNN
processing. Results are generated with 6,000 runs.}
}  
\label{table:4}
\centering
\begin{tabular}{| c || c | c  | c | c | c |} 
 \hline
   &   \multicolumn{5}{c|}{Background Pixel Value}  \\ 
 \hline
  Board &0 (default)&1&10&30&50\\
  \hline
  \hline
 ZCU104 (w/o denoising)& 0.75 & 0.46 & 0.53  & 0.56 & 0.56 \\
 ZCU104 (w/ denoising)& 0.82 & 0.62 & 0.65  & 0.67 & 0.66 \\
 VCU118 (w/o denoising)  & 0.71 &  0.59 & 0.59 & 0.60 & 0.58 \\
 VCU118 (w/ denoising) & 0.76 &  0.70 & 0.63 & 0.64 & 0.64 \\

 \hline
\end{tabular}
\end{table}

\begin{figure}[t]
	\centering
    \includegraphics[width=0.8\linewidth]{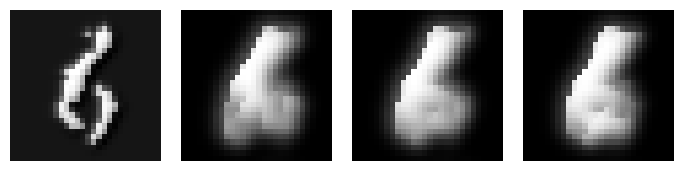}
 \caption{\small {\color{black} Recovered images from ZCU104 board when background
pixels are replaced with non-zero values, after applying the denoising
algorithm. The background values are 0 (default), 1, 30, and 50 from
left to right. Each experiment performed with 6,000 runs of the
same modified image.}} 
\label{fig:example_background}
\end{figure}

For further experimentation, we created groups of 6,000 distinct images
of Figure~\ref{fig:sixes1} by flipping the least significant bit (LSB) or two least
significant bits of each input pixel with a  fixed probability. 
Effectively,
this process mimics 
analog-to-digital converter noise that may be present during image
sampling.
Then,
we collected TDC measurements from each noisy image once and used the
mean of all the 6,000 traces to recover the input image.

 Table \ref{table:5} shows the normalized cross-correlation of the experiments
for six different bit flipping probabilities (0 indicates that no bits
were flipped and 100 indicates that all LSB or least significant two bits were flipped). In cases of
two-bit flips, both bits of the pixel were flipped from their original
values. The results show that bit flipping has a limited effect on
normalized cross correlation, although always flipping the LSBs does
show some degradation. Sample reconstructed images shown in Figure \ref{fig:example_flip_all} indicate continued visual recognition. 

\begin{table}[t]
\caption{{\color{black} Normalized cross-correlation of the denoised recovered
image and the original image shown in  Figure~\ref{fig:sixes1} if the
least significant bit or least significant two bits of all pixels are
flipped prior to input to the BNN accelerator. 
Results are generated with 6,000 runs.}}

\label{table:5}
\centering
\begin{tabular}{| c || c | c  | c | c | c | c |} 
 \hline
   &   \multicolumn{6}{c|}{Flipping Probability}  \\ 
 \hline
  Board (Bit Number)&0&20&40&60&80&100\\
  \hline
  \hline
 ZCU104 (LSB)& 0.83 & 0.77 & 0.80  & 0.75 & 0.76 & 0.63 \\
 VCU118 (LSB)& 0.76 & 0.76 & 0.73 & 0.67 & 0.67 & 0.64 \\
 ZCU104 (2 Bits) & 0.83 & 0.76 & 0.76  & 0.73 & 0.74 & 0.62\\
 VCU118 (2 Bits) & 0.76 &  0.75 & 0.70 & 0.63 & 0.65 & 0.63 \\

 \hline
\end {tabular}
\end{table}

\begin{figure}[t]
	\centering
	\begin{subfigure}{1\linewidth}
	    \centering
	    \captionsetup{justification=centering}
	    \includegraphics[width=0.8\linewidth]{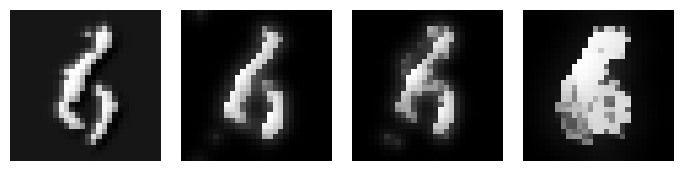}
	    \caption{\small Flip LSB, Probability from left to right = 0, 40, 80, 100\% }
	    \label{fig:example_flip}
	\end{subfigure}
    \begin{subfigure}{1\linewidth}
    	\centering
	\captionsetup{justification=centering}
        \includegraphics[width=0.8\linewidth]{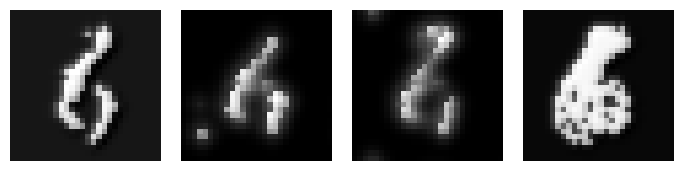}
    	\caption{\small Flip lower 2 bits, Probability from left to right = 0, 40, 80, 100\% }
    	\label{fig:example_flip2bit}
    \end{subfigure}	
\caption{\small {\color{black} Recovered images from the ZCU104 board for 6,000 runs:
(a) denoised
recovered images after flipping the least significant bit of pixels in the BNN input image with fixed
probabilities, (b) denoised recovered images
for probabilistic flips of the bottom two pixel bits.}}

\label{fig:example_flip_all}
\end{figure}

In a final experiment to explore limitations, we evaluated the
reconstruction of several images from the Fashion MNIST dataset \cite{fashion-mnist} of 
black-and-white garment images with the same input image size of 28$\times$28 pixels as the MNIST handwritten digit database. These
images of garments have a broader range of textures than digits. As seen in Figure
\ref{fig:example_fashion}, image reconstruction of a sample of input
images shows a distinct recognizable garment outline although internal
garment textures are missing.

\begin{figure}[t]
	\centering
	\begin{subfigure}{0.45\textwidth}
	    \centering
	    \captionsetup{justification=centering}
	    \includegraphics[width=0.8\linewidth]{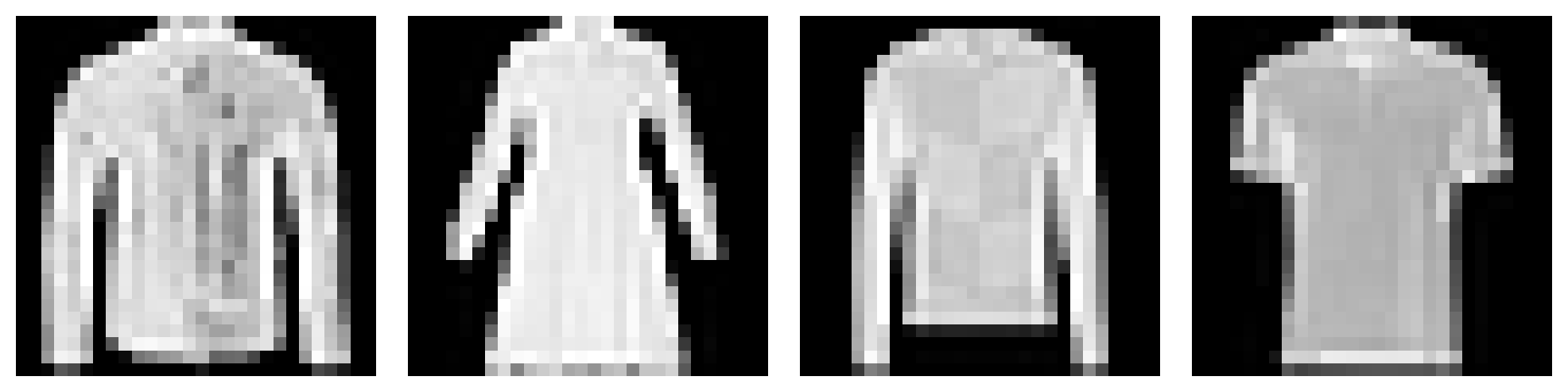}
	    \caption{\small Input Image, left to right: coat, dress, pullover, T-shirt}
	    \label{fig:example_fashion1}
	\end{subfigure}
    \begin{subfigure}{0.45\textwidth}
    	\centering
	\captionsetup{justification=centering}
        \includegraphics[width=0.8\linewidth]{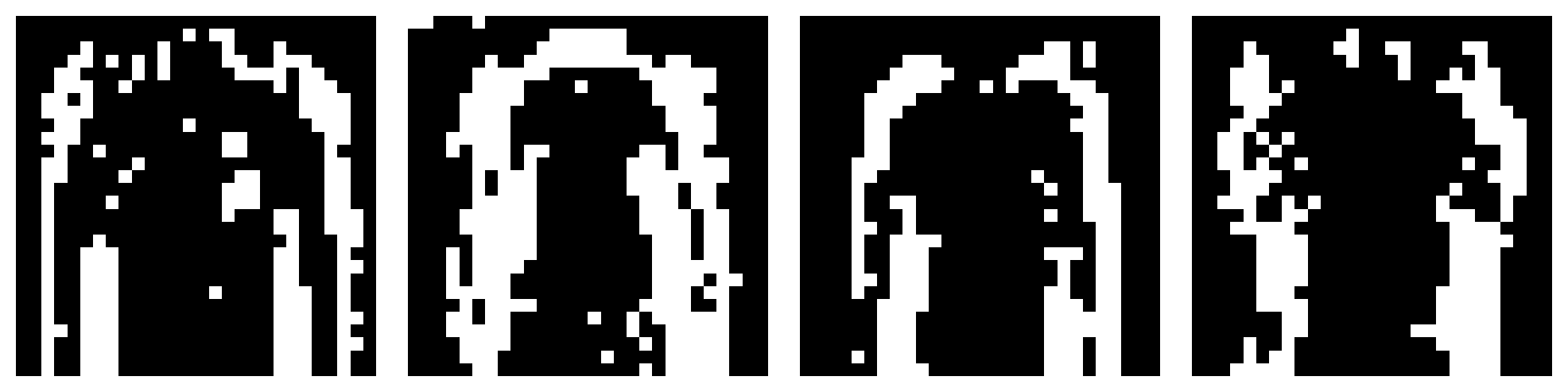}
    	\caption{\small Recovered images using the ZCU104 board}
    	\label{fig:example_fashion2}
    \end{subfigure}	
    \begin{subfigure}{0.45\textwidth}
        \centering
        \captionsetup{justification=centering}
        \includegraphics[width=0.8\linewidth]{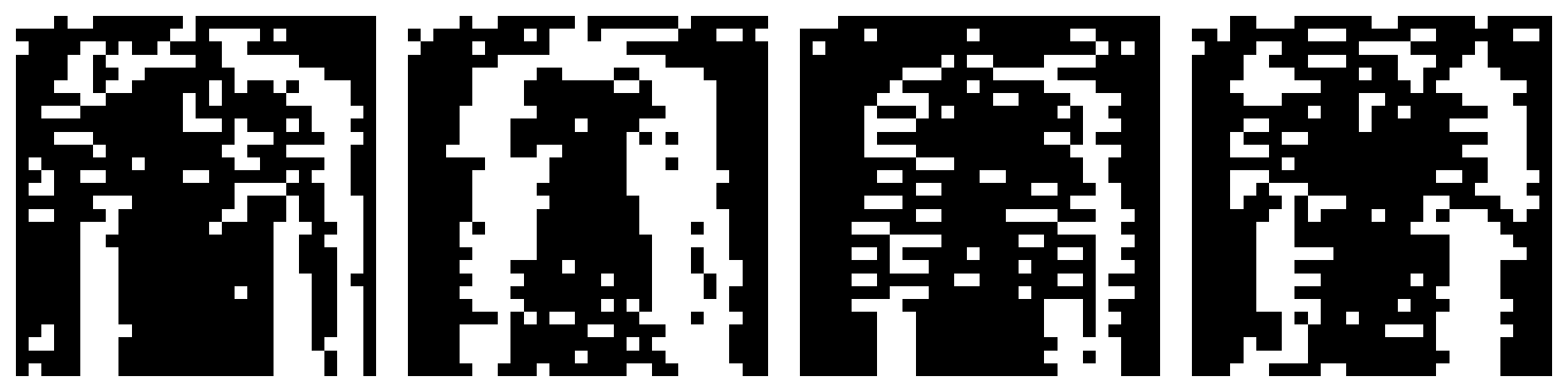}
    	\caption{\small Recovered images using the VCU118 board}
    	\label{fig:example_fashion3}
    \end{subfigure}	\\
 \caption{\small {\color{black} Recovered images for coat, dress, pullover, and T-shirt images from the Fashion-MNIST dataset using
the ZCU104 and VCU118 boards for 6,000 runs.}} 	
\label{fig:example_fashion}
\end{figure}

}

\section{conclusion}
\label{sec_conclustion}
This paper presents a
remote power side-channel attack on binarized convolutional neural networks
targeting multi-tenant FPGAs. We show 
that it is possible to accurately extract image inputs to a BNN accelerator
by collecting and analyzing on-chip
voltage estimates. 
Time-to-digital converters are leveraged to obtain
voltage estimates on the FPGA chip 
during execution of the algorithm. 
Our approach has been successfully applied to four FPGA boards,
including on Xilinx UltraScale+ FPGAs located on Amazon AWS F1 cloud
servers. Our experiments successfully recovered recognizable images for
all ten digits from the MNIST handwritten digit database.

{\color{black} This research opens up significant avenues for future exploration.
Additional attacks to extract kernel values are needed to identify both
BNN image inputs and  parameters. The collection and analysis of
voltage estimates for multi-kernel processing in the first convolution
layer could be used in this effort. Additional layers of the BNN may
also be vulnerable to the extraction of voltage estimates. It may also
be possible to use a similar approach to extract input images for CNNs
with non-binary kernel values. Such an approach would require the use
of multipliers for weight scaling, possibly leading to increased power
consumption. More complex datasets, including color
images, and applications, such as face recognition, could also be considered. 
More research is also needed to determine exactly when BNN
processing starts so that TDC sampling can be synchronized with 
BNN processing.

Countermeasures are also needed to reduce the effectiveness of on-chip
voltage measurement attacks. The extraction of voltage estimates could
be impeded by the significant circuit switching of interfaces or other
design components in the proximity of the convolution unit (e.g.,
active fences~\cite{krautter2019active}). 
Additionally, the pixel order of convolution unit processing could be
scrambled on a per-image basis to make image reconstruction more difficult.}

\bibliographystyle{IEEEtran}
\bibliography{main}

\vspace{-0.5in}
\begin{IEEEbiography}[{\includegraphics[width=1in,height=1.25in,clip,keepaspectratio]{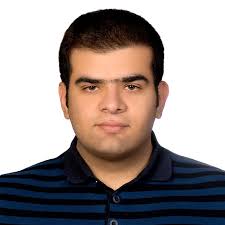}}]
{Shayan Moini} (S16) received his B.Sc. degree In
Electrical Engineering at
Sharif University of Technology, Tehran, Iran, in 2014. In 2017, he
completed his M.Sc. degree in Electrical Engineering 
at the University of Tehran, Iran. He is currently studying towards a Ph.D. degree
in the Department of Electrical and Computer Engineering at the
University of Massachusetts, Amherst, MA, USA. 

\end{IEEEbiography} 

\vspace{-0.6in}
\begin{IEEEbiography}[{\includegraphics[width=1in,height=1.25in,clip,keepaspectratio]{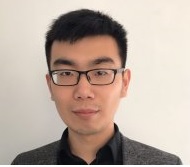}}]
{Shanquan Tian} (S18) received his B.S. degree in Applied Physics from
University of Science and Technology of China, Anhui, China in 2017,
and the M.S. degree in Electrical Engineering from Yale University, New
Haven, CT, USA in 2019, where he is currently studying towards a Ph.D.
degree in Computer Architecture and Security Laboratory, Department of
Electrical Engineering at Yale. 
\end{IEEEbiography} 

\vspace{-0.6in}
\begin{IEEEbiography}[{\includegraphics[width=1in,height=1.25in,clip,keepaspectratio]{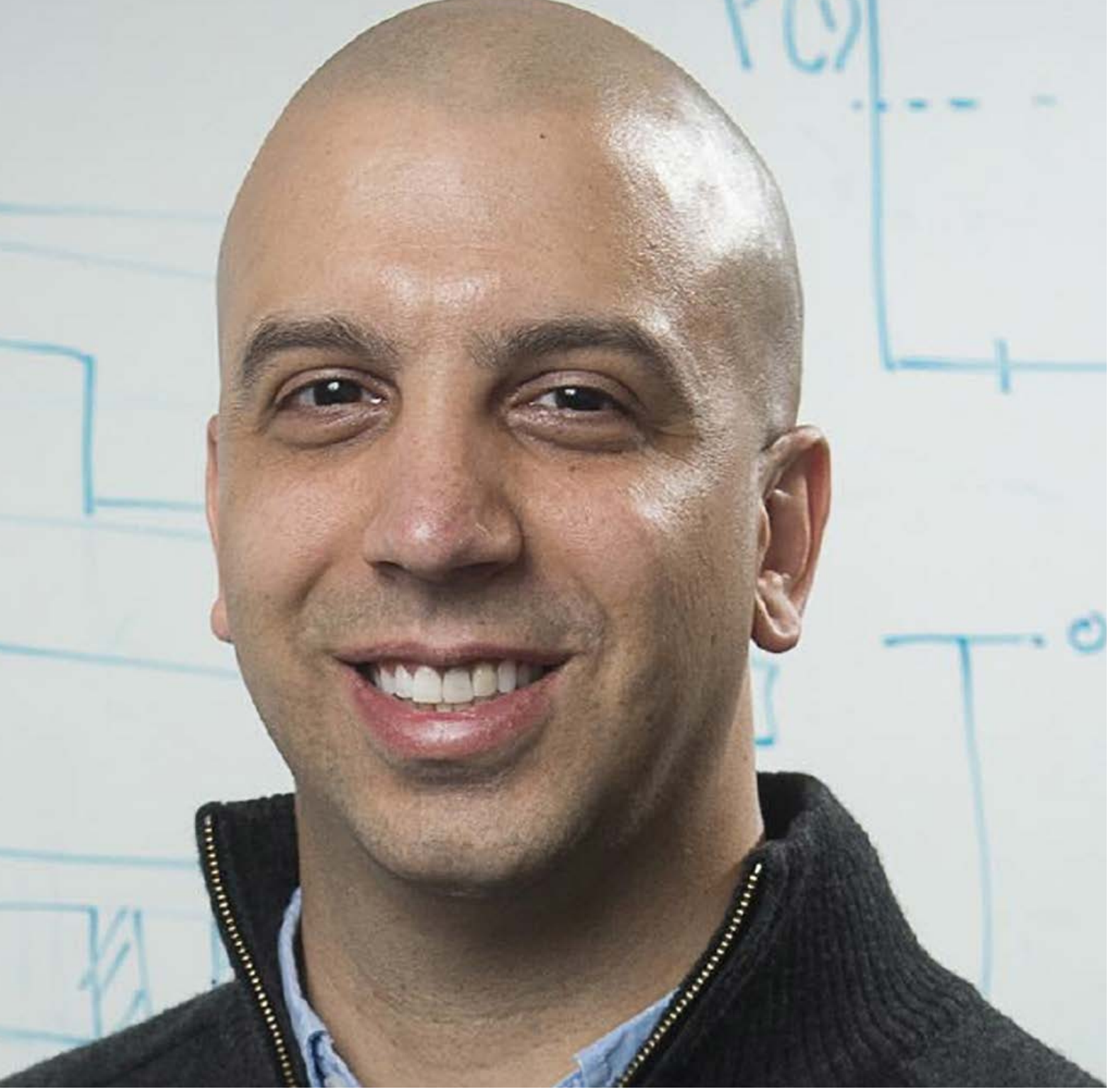}}]
{Daniel Holcomb} (M07) received the B.S. and M.S. degrees in electrical
and computer engineering from the University of Massachusetts Amherst,
Amherst, MA, USA, and the Ph.D. degree in electrical engineering and
computer sciences from the University of California at Berkeley,
Berkeley, CA, USA. He is currently an Associate Professor with the
Department of Electrical and Computer Engineering, University of
Massachusetts Amherst. 
\end{IEEEbiography} 

\vspace{-0.6in}
\begin{IEEEbiography}[{\includegraphics[width=1in,height=1.25in,clip,keepaspectratio]{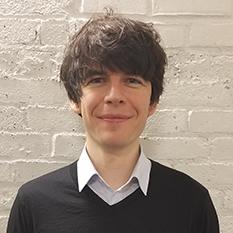}}]
{Jakub Szefer} (S08--M13--SM19) received B.S. with highest honors in Electrical and
Computer Engineering from University of Illinois at Urbana-Champaign,
and  M.A. and Ph.D. degrees in Electrical Engineering where his research
focused on secure hardware architectures.  He is currently an Associate
Professor of Electrical Engineering at Yale University where he leads
the Computer Architecture and Security Laboratory (CASLAB).  
\end{IEEEbiography}

\vspace{-0.6in}
\begin{IEEEbiography}[{\includegraphics[width=1in,height=1.25in,clip,keepaspectratio]{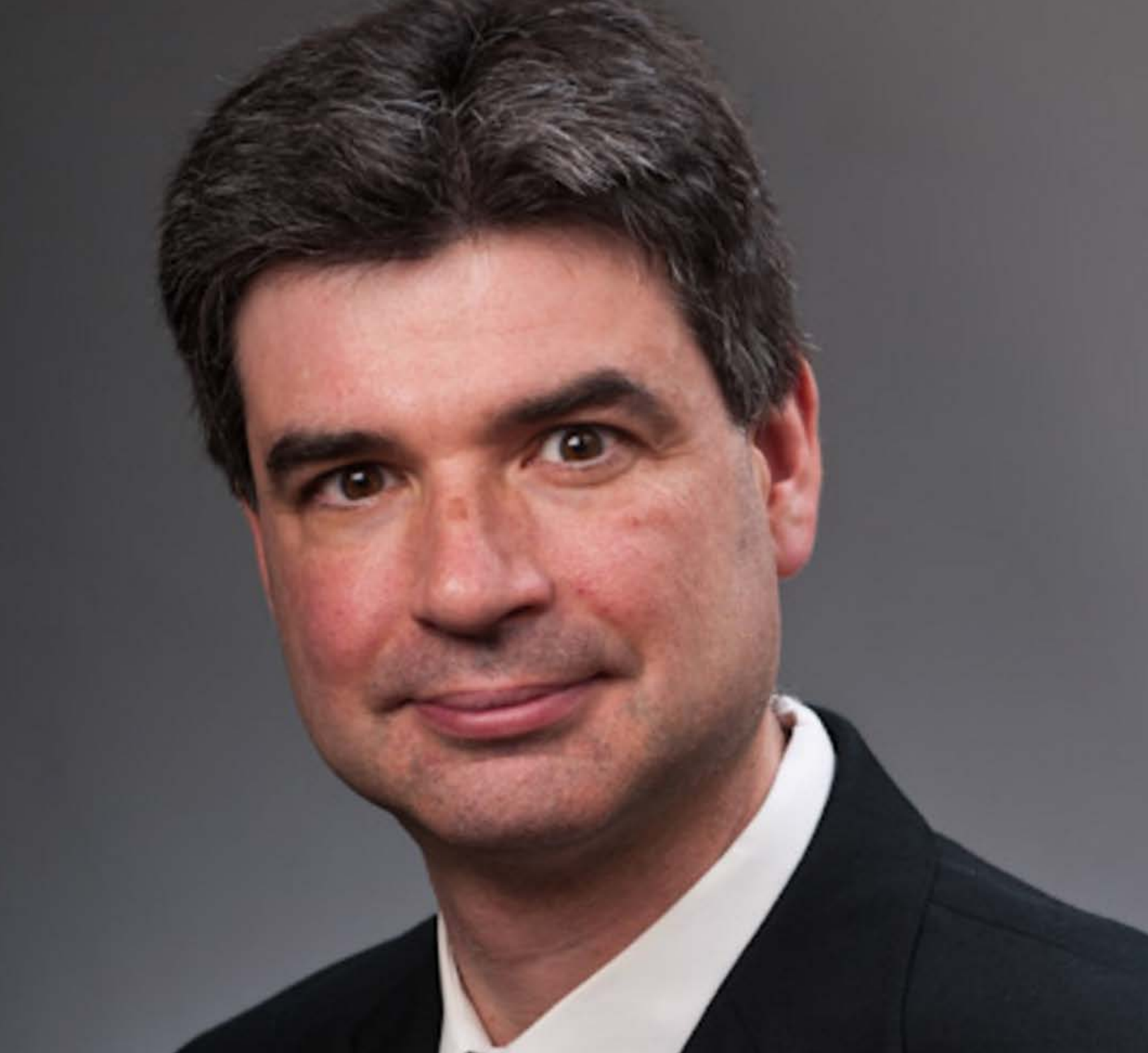}}]
{Russell Tessier} (M00--SM07) received the B.S. degree in computer and
systems engineering from Rensselaer Polytechnic Institute, Troy, NY,
USA and the S.M. and Ph.D. degrees in electrical engineering
from the Massachusetts Institute of Technology, Cambridge, MA, USA.
He is currently Professor of Electrical
and Computer Engineering with the University of Massachusetts, Amherst,
MA. 
\end{IEEEbiography}

\end{document}